\documentstyle[12pt,a4,psfig,epsfig]{article}
\newcommand{\vs}[1]{\rule[- #1 mm]{0mm}{#1 mm}}

\begin{document}
\renewcommand{\thefootnote}{\fnsymbol{footnote}}
\newpage
\pagestyle{empty}
\setcounter{page}{0}

\null
\begin{center}

{\Large {\bf PHOTOPRODUCTION OFF NUCLEI}}\\[0,3cm]
{\Large {\bf AND POINT-LIKE PHOTON INTERACTIONS}}\\[0,5cm]
{\Large {\bf Part I: Cross Sections and Nuclear Shadowing}}

\vs{5}

{\bf R.\ Engel}\\
{\em Universit\"at Leipzig, Fachbereich Physik, D-04109 Leipzig, Germany
\\ and Universit\"at Siegen, Fachbereich Physik, D-57068 Siegen, Germany}

\vs{0,5}

{\bf J.\ Ranft}\\
{\em Departamento de F\'\i sica de Part\'\i culas, Universidade de Santiago
     de Compostela,\\
     E-15706 Santiago de Compostela, Spain}

\vs{0,5}

{\em and}

\vs{0,5}

{\bf S.\ Roesler}\\
{\em Universit\"at Siegen, Fachbereich Physik, D-57068 Siegen, Germany}

\end{center}
\vs{5}

\centerline{ {\bf Abstract}}
\indent

High energy photoproduction off nuclear targets is studied within 
the Glauber-Gribov approximation.
The photon is assumed to interact as a $q\bar{q}$-system according to the
Generalized Vector Dominance Model and as a ``bare photon'' in direct
scattering processes with target nucleons. We calculate total cross 
sections for interactions of photons with nuclei 
taking into account coherence length effects and point-like interactions 
of the photon. Results are compared to data on photon-nucleus
cross sections, nuclear shadowing, and quasi-elastic $\rho$-production.
Extrapolations of cross sections and of the shadowing behaviour to high 
energies are given.

\vfill
\rightline{Siegen SI 96-09}
\rightline{Santiago de Compostela US-FT/42-96}
\rightline{October, 1996}

\newpage
\pagestyle{plain}
\renewcommand{\thefootnote}{\arabic{footnote}}
%
%
\section {Introduction}
During the last years, the understanding of photon-hadron interactions
has considerably improved due to new experimental data from photoproduction
and low-$x$ measurements in $ep$ collisions at HERA and due to their
interpretation in terms of QCD-inspired multiple-interaction 
models~\cite{Erdmann96}. Experimental evidence for the classification of 
photon interactions within the parton model into {\it direct} and 
{\it resolved} interactions has been found~\cite{Ahmed92a,Derrick94a}. 
Both classes of processes show different features concerning 
cross sections as well as multiparticle production~\cite{Aid95c,Derrick95f}.

Within the QCD-improved parton model, in direct processes the photon couples 
directly to a parton of the hadron, whereas in resolved processes it
enters the scattering process as a hadronic quark-antiquark fluctuation.
Resolved processes are well described in the framework of the Generalized 
Vector Dominance Model (GVDM) (see for example~\cite{Bauer78,Donnachie78} 
and references therein).
Due to relatively large lifetimes the $q\bar q$-states may develop
properties of ordinary hadrons by emitting and absorbing virtual
partons. Therefore, the $q\bar q$-states can either interact with the hadron
in soft scattering processes (the produced final-state particles have
small transverse momenta) or the partons of the $q\bar q$-system can
participate in hard interactions with the partons of the
hadron~\cite{Engel95a,Engel95d}.
In case of a hard interaction and if the mass of the $q\bar q$-fluctuation 
is large in comparison to the perturbative QCD
scale $\Lambda_{\mbox{\scriptsize QCD}}$, it is possible to calculate
not only the hard parton-parton scattering but also the splitting of the
photon into the $q\bar q$-pair perturbatively. These hard resolved
interactions of high-mass $q\bar q$-states are frequently called {\it
anomalous} photon interactions~\cite{Witten77}.

Here, we want to study the implications of the above mentioned experimental 
findings on direct and resolved processes to the understanding of 
photon-nucleus collisions at comparable or higher photon energies.

High energy
photon-nucleus collisions have been studied experimentally and theoretically 
by numerous groups (for recent reviews we refer 
to~\cite{Arneodo94b,Piller95}). 
It was found that they bear a remarkable 
resemblance to hadron-nucleus interactions. For instance, both show decreasing
per-nucleon cross sections with increasing nucleus mass number, an effect 
which is known as ``shadowing''. Again, the GVDM provides a natural 
interpretation of these photon-hadron similarities~\cite{Bauer78,Donnachie78}. 
Like in hadron-nucleus collisions, shadowing in photon-nucleus collisions 
can then be described in the framework of the Gribov-Glauber 
approximation~\cite{Glauber55,Gribov70,Bertocchi72}. It relates the total
photon-nucleus cross section to effective $q\bar q$-nucleon cross 
sections~\cite{Piller95,Frankfurt89,Frankfurt94,Davidenko78,Ioffe84a}. 
However, the application of 
the Gribov-Glauber formalism to the multiple scattering process of a
$q\bar q$-state without further constraints is only justified if the 
interaction length exceeds the nuclear 
radius~\cite{Huefner96a}. This is not the case for the above mentioned direct
processes since the photon interacts in such processes with only one nucleon.
As we will argue further below, at high energies it might also be not the 
case for anomalous photon interactions. In particular, we will consider
the extreme assumption that the interaction time is less than any internucleon
distance, i.e.\ that again only one target nucleon is involved.
For these reasons, in the following we call direct and anomalous 
photon interactions {\it point-like} processes. They may lead to a suppression
of the Glauber-multiple scattering process, i.e. to a suppression of 
shadowing. It can be expected that this feature is most clearly pronounced 
in photon scattering processes off heavy nuclei and at high energies, where 
the cross section of the point-like photon-nucleon interaction becomes 
sizeable as compared to the total photon-nucleon cross section.

The intention of the present paper is twofold: calculating cross
sections of photon-nucleus interactions we (i) investigate the influence of
point-like processes on the shadowing behaviour at high energies and
(ii) provide the basis for a forthcoming study of particle production in
photon-nucleus collisions~\cite{Engel96c}. 
In Sect.~2 we consider photon-nucleon collisions and
derive total cross sections for $q\bar q$-nucleon interactions. In Sect.~3
point-like photon interactions are discussed and their contribution to
the total photon-nucleon cross section is estimated. In Sect.~4 we
calculate photon-nucleus cross sections and the shadowing behaviour.
Both are compared to data on photoproduction and Deep Inelastic
Scattering (DIS) off nuclei and extrapolations to high energies are given.
Finally, in Sect.~5 we summarize our results.
\noindent
%
%
\section{\label{Sect_sig-gn}
         Total photon-nucleon cross sections}
Throughout this paper we consider the photon-nucleon scattering process
in the laboratory-frame (nucleon rest frame) using the following 
kinematical variables.
The Bjorken-$x$ variable is defined as $x=Q^2/2m\nu$ denoting with 
$Q^2$, $\nu$, and $m$ the photon virtuality, the photon energy,
and the nucleon mass, resp. The squared total energy of the
photon-nucleon system is given by $s=Q^2(1-x)/x+m^2$. We restrict our 
discussions to small $x$-values ($x<0.1$) and to the limit $s\gg Q^2$.

Within the diagonal GVDM~\cite{Bauer78,Donnachie78} it is assumed 
that the virtual photon fluctuates into intermediate $q\bar{q}$-states $V$
of mass $M$ which subsequently may interact with the nucleon $N$.
This fact can be expressed by a spectral relation of the
form~\cite{Donnachie78,Piller95,Sakurai72a}
\begin{equation}
\label{GVDM-sig_gn}
\sigma_{\gamma^{\star}N}(s,Q^2)=4\pi\alpha_{\mbox{\scriptsize em}} 
\int_{M_0^2}^{M_1^2} dM^2\ D(M^2) \left(\frac{M^2}{M^2+Q^2}\right)^2
\left(1+\epsilon\frac{Q^2}{M^2}\right) \sigma_{VN}(s,Q^2,M^2).
\end{equation}
We use $\alpha_{\mbox{\scriptsize em}}=e^2/4\pi=1/137$. 
The factor $D(M^2)$ incorporates the density of $q\bar{q}$-systems per
unit mass-squared interval:
\begin{equation}
\label{GVDM-sig_gn-D}
D(M^2)=\frac{R_{e^+e^-}(M^2)}{12\pi^2M^2}, \qquad 
R_{e^+e^-}(M^2)=\frac{\sigma_{e^+e^-\longrightarrow \mbox{\scriptsize
hadrons}}(M^2)}{\sigma_{e^+e^-\longrightarrow \mu^+\mu^-}(M^2)}\approx
3 \sum_f e_f^2,
\end{equation}
where we sum up the squared quark charges of all quark flavors being
energetically accessible.
$\epsilon$ is the ratio between the fluxes of longitudinally and
transversally polarized photons.
$\sigma_{VN}$ denotes the effective cross section for the interaction of 
a $q\bar{q}$-system with mass $M$ with a nucleon.

Considering low-$Q^2$
$\gamma^\star p$ scattering only, a detailed model for the $M^2$- and
$Q^2$-dependence of $\sigma_{VN}$ is not needed. At high collision
energies, the average lifetime of the hadronic $q\bar q$-fluctuation
$t_f\sim 2\nu/(M^2+Q^2)$ is almost always larger than the typical hadronic 
interaction time $t_{\mbox{\scriptsize int}}$ (for nucleons 
$t_{\mbox{\scriptsize int}}\sim r_N\approx 1$~fm).
However, in photon-nucleus collisions the $M^2$- and $Q^2$-dependence 
of $\sigma_{VN}$ might be important since the coherence length $d\sim
t_f$ of the 
hadronic fluctuation can become comparable to or
smaller than the nuclear radius or the nuclear mean free path ($\approx
1/(n\sigma_{VN}$), with $n$ being the number of nucleons per unit 
volume)~\cite{Donnachie78}.
Therefore, in the following we are going to estimate the (purely
theoretical) quantity $\sigma_{VN}$ using Eq.(\ref{GVDM-sig_gn}) and a
parametrization for the experimentally measurable cross section
$\sigma_{\gamma^\star N}$.

With increasing mass $M$ of the $q\bar{q}$-system the virtuality of 
the $q$ and $\bar q$ of the system increases.
As a consequence the transverse size of the hadronic
fluctuation and, hence, $\sigma_{VN}$
decreases like $1/M^2$ at large $M^2$~\cite{Donnachie78,Ioffe84a}.
Following Ref.~\cite{Donnachie78}
we approximate this effect parametrizing $\sigma_{VN}$ as
\begin{equation}
\label{GVDM-sig_gn-sig_VN}
\sigma_{VN}(s,Q^2,M^2)=\frac{\tilde{\sigma}_{VN}(s,Q^2)}
{M^2+Q^2+C^2}.
\end{equation}
Here, $C$ is a model-dependent parameter~\cite{Donnachie78} and
taken to be $C^2=2$~GeV$^2$. 
With Eq.(\ref{GVDM-sig_gn-sig_VN}) the $M^2$-dependence of the integrand 
in Eq.(\ref{GVDM-sig_gn}) is explicitly
known and the integration over $M^2$ between $M_0^2=4m_{\pi}^2$ and
$M_1^2=s$ can be performed. The lower integration limit
corresponds to the kinematical threshold. Alternatively, the contributions
from the low mass vector mesons $\rho^0$, $\omega$, and $\phi$ could be added
as separate terms to the continuum (Eq.(\ref{GVDM-sig_gn})), in this case 
starting the integration at $m_{\phi}^2$~\cite{Piller95,Gorczyca73}.
However, this has been omitted for simplicity. The upper
limit, here formally taken to be $s$, has practically no influence on
the results at low and moderate $Q^2$ since high $M^2$-values are
suppressed. The only quantity on the r.h.s. of
Eqs.(\ref{GVDM-sig_gn},\ref{GVDM-sig_gn-sig_VN})
which is unknown so far is $\tilde{\sigma}_{VN}$. 
Using a parametrization for $\sigma_{\gamma^{\star}N}$, the $M^2$-independent
part of Eq.(\ref{GVDM-sig_gn-sig_VN}), $\tilde{\sigma}_{VN}$, can be 
calculated for each value of $s$ and $Q^2$.

Applying the convention of Ref.~\cite{Budnev75}, the cross section for
the scattering of virtual photons off nucleons $\sigma_{\gamma^{\star}N}$ 
can be written as
\begin{equation}
\label{sigF2}
\sigma_{\gamma^{\star}N}(s,Q^2)=\frac{4\pi^2\alpha_{\mbox{\scriptsize
em}}}{Q^2(1-x)}F_2^N(x,Q^2)\ .
\end{equation}
Since in the present paper we want to study cross sections and shadowing
in the $Q^2$-region of both, photoproduction and DIS, we use the model of
Capella {\it et al.}~\cite{Capella94b} (CKMT model) for
the structure function $F_2^N$ which provides a simple analytical
parametrization valid for $0 \le Q^2 \stackrel{<}{\sim} 5$~GeV$^2$. 
The nucleon structure function is derived from Regge
arguments taking rescattering effects into account:
\begin{equation}
\label{f2deut}
F_2^N(x,Q^2)=A x^{-\Delta(Q^2)} (1-x)^{n(Q^2)+4} \left(
\frac{Q^2}{Q^2+a}\right)^{1+\Delta(Q^2)}+B\ x^{1-\alpha_R} (1-x)^{n(Q^2)} 
\left(\frac{Q^2}{Q^2+b}\right)^{\alpha_R}
\end{equation}
with
\begin{equation}
\Delta(Q^2) = \Delta_0\left(1+\frac{2Q^2}{Q^2+d}\right), \qquad
n(Q^2) = \frac{3}{2} \left(1+\frac{Q^2}{Q^2+c}\right).
\end{equation}                                                           
The first term in~(\ref{f2deut}) is
associated with the pomeron contribution determining the small-$x$
behaviour of the sea-quark distribution function whereas the second 
term is parametrized according to secondary reggeon contributions governing 
the valence-quark distribution function of the nucleon.
We refer to~\cite{Capella94b} for the values of the parameters entering the 
expressions. The structure function resulting from this model is in 
reasonable agreement with measurements~\cite{Capella94b}.
Using the ansatz for the gluon distribution as given 
in~\cite{Capella94b} we obtain $F_2^N$ for higher values of $Q^2$ 
by performing a QCD evolution in leading logarithmic approximation.

We note that also other parametrizations would be suitable, for instance
the parametrization of Abramowicz {\it et al.}~\cite{Abramowicz91a}
and in the low $Q^2$-range the parametrizations 
of Badelek and Kwieci\'nski~\cite{Badelek92}.
Furthermore, for low values of $Q^2$ the
photon-nucleon cross section $\sigma_{\gamma^{\star}N}$ can be equally 
well obtained in the framework of the two-component Dual Parton 
Model (DPM)~\cite{Engel95a,Engel95d}. 
The advantage of the two-component DPM calculation is that in this case 
a detailed model for the inelastic final states exists~\cite{Engel95d},
which we will also apply to the study of particle production in a forthcoming
paper~\cite{Engel96c}.

In Fig.~\ref{gptot} we compare the photoproduction cross sections 
$\sigma_{\gamma p}^{\mbox{\scriptsize tot}}$ obtained from the CKMT-model 
(thick solid line) and calculated within the two-component DPM (dotted line) 
with data~\cite{Alekhin87,Derrick94b,Aid95b}. The differences in the high
energy extrapolation reflect the typical size of the theoretical 
uncertainties.

In Fig.~\ref{gpsigvn}~a) we show the energy-dependence of the effective
cross section $\sigma_{VN}$ for $M^2 = m_\rho^2$ and
different $Q^2$ values. As observed in $\gamma^\star p$ collisions, the
rise of the cross section with energy becomes steeper with
increasing photon virtuality.
In Fig.~\ref{gpsigvn}~b) the $Q^2$-dependence of $\sigma_{VN}$ for different
energies and $M^2 = m_\rho^2$ is given. As expected,
the $Q^2$ dependence is very weak for $Q^2<m_\rho^2+C^2$.
\section{\label{Sect_sig-pl}
         Contributions from point-like interactions of the photon}
It should be emphasized that resolved as well as direct photon
interactions are included in the description of photon-nucleon
scattering via the GVDM (Eq.(\ref{GVDM-sig_gn})). 
Of course, a sharp distinction between direct 
and resolved interactions is not possible. 
In direct interactions, the photon couples directly to a parton of the 
nucleon which determines the highest virtuality of the scattering process 
(see Fig.~\ref{grf_plike}~a). In resolved interactions, the photon may
fluctuate into a $q\bar q$-pair with high virtuality. 
For example, a $q\bar q$-system can emit a gluon leading to a quark with
even higher virtuality which couples to a gluon of the nucleon 
as shown in Fig.~\ref{grf_plike}~b).
Therefore, it is necessary to distinguish two scales to characterize a hard 
photon-nucleon scattering: 
(i) the virtuality $M$ of the hadronic $q\bar q$-fluctuation
and (ii) the scale of the hard scattering $\mu$ which is approximately
given by the momentum transfer in the hard scattering process.
Then, for $\mu^2 \approx  M^2$ the interaction is classified as direct
interaction whereas for $\mu^2 \gg M^2$ the interaction is a resolved one.
As already mentioned, in case of resolved interactions with
$\mu^2 \gg M^2$ and $M^2 \gg \Lambda_{\mbox{\scriptsize QCD}}^2$, not only
the hard parton-parton scattering but also the splitting of the photon
into the $q\bar q$-pair can be calculated perturbatively. These
interactions lead to a
rise of the photon structure function with $\mu^2$ like $\ln(\mu^2)$
(anomalous contribution to the photon structure function~\cite{Witten77}).
In the following we consider as point-like photon interactions all processes
which are characterized by 
$M^2 \gg \Lambda_{\mbox{\scriptsize QCD}}^2$~\cite{Schuler93b}.

In direct and anomalous processes either $M^2$ or the transverse momentum 
of the hard scattering, $p_\perp$, acts as hard scale permitting 
perturbative calculations. Therefore, the cross section for direct processes 
$\sigma_{\gamma^{\star}N}^{\mbox{\scriptsize dir}}$ and 
the cross section $\sigma_{\gamma^\star N}^{\mbox{\scriptsize ano}}$ for 
the fluctuation of a photon into a $q\bar q$-system with a large mass $M$ 
(i.e.\ highly virtual quarks) and the interaction of this system with a 
nucleon can be estimated using perturbative QCD.

In lowest-order perturbative QCD, the direct photon-nucleon cross
section follows from
\begin{equation}
\sigma_{\gamma^{\star}N}^{\mbox{\scriptsize dir}}
(s,p_\perp^{\mbox{\scriptsize cutoff}}) =
\int dx d\hat{t}\ \sum_{i,k,l}
f_{i|N}(x,\mu^2) \frac{d\sigma_{\gamma,i\rightarrow
k,l}^{\mbox{\scriptsize QCD}}(\hat{s},\hat{t})}{d\hat{t}}
\Theta(p_\perp-p_\perp^{\mbox{\scriptsize cutoff}}),
\label{hard-dir}
\end{equation}
where $f_{i|N}$ denotes the parton distribution function (PDF)
for the parton $i$ of the nucleon and the sum runs over all possible
parton configurations $(i,k,l)$. For the calculation we use $\mu^2 =
p_\perp^2/4$. The transverse momentum cutoff $p_\perp^{\mbox{\scriptsize
cutoff}}$ restricts the integration to the perturbatively reliable region.

In order to calculate the anomalous cross section
$\sigma_{\gamma^{\star}N}^{\mbox{\scriptsize ano}}$, we use the {\sc
Phojet} Monte Carlo (MC) event generator~\cite{Engel95a,Engel95d}
to simulate hard resolved photon-nucleon interactions according to
the cross section
\begin{eqnarray}
\sigma^{\mbox{\scriptsize res}}_{\gamma N}(s,p_\perp^{\mbox{\scriptsize 
cutoff}}) = \int dx_1 dx_2
d\hat{t} \sum_{i,j,k,l} &\bigg(&\!\!\!\frac{1}{1+\delta_{k,l}}
f_{i|\gamma}(x_1,\mu^2) f_{j|N}(x_2,\mu^2) \nonumber\\
&\times&\frac{d\sigma_{i,j\rightarrow
k,l}^{\mbox{\scriptsize QCD}}(\hat{s},\hat{t})}{d\hat{t}}
\Theta(p_\perp-p_\perp^{\mbox{\scriptsize cutoff}}) \bigg)\ .
\label{cs-hard}
\end{eqnarray}
This cross section receives contributions from low-mass and high-mass
$q\bar q$-fluctuations. In order to determine the cross section due to 
anomalous interactions, initial state parton showers were generated for each 
hard interaction  using a backwards evolution algorithm similar to the one 
discussed in~\cite{Sjostrand85,Gottschalk86}
using the parton transverse momentum as evolution variable. 
Some basic coherence effects are implemented by imposing
angular ordering of the parton emissions.
Furthermore, the possibility to have a hard $\gamma \rightarrow q \bar q$
process during the shower evolution is taken into account.
After each parton emission, the probability to stop the parton shower
evolution due to a point-like splitting is taken to be the ratio
of the $\gamma \rightarrow q \bar q$ contribution to the quark density in 
the photon
\begin{equation}
q(x,\mu^2) = \frac{3 \alpha_{\mbox{\scriptsize em}}}{2 \pi} e_q^2
\left[ \left(x^2+(1-x)^2\right)
\ln\left(\frac{1-x}{x}\frac{\mu^2}{(p_\perp^{\mbox{\scriptsize cutoff}})^2}
\right) + 8 x (1-x) - 1
\right]
\label{ano-cont}
\end{equation}
and the quark density of the full photon PDF. 
Since we are only interested in $\gamma \rightarrow q \bar q$ splittings
with a remnant quark having $p_\perp > p_\perp^{\mbox{\scriptsize cutoff}}$, 
the transverse momentum cutoff $p_\perp^{\mbox{\scriptsize cutoff}}$ is
used in (\ref{ano-cont}) as the lowest quark virtuality.
Then, the fraction of the anomalous cross section to the total hard
resolved photon-nucleon cross section is given by the fraction of
events where an anomalous splitting with $p_\perp>p_\perp^{\mbox{\scriptsize
cutoff}}$ has been generated.
Within the calculations, we use the GRV PDF parametrizations for the 
photon~\cite{GRV92b} and the proton~\cite{GRV92a}.
For the transverse momentum cutoff a value of 3~GeV/$c$ is applied 
consistently to both, direct and resolved interactions.
Regarding the choice of this value we refer to a forthcoming 
paper~\cite{Engel96c}.
In this paper we want to study particle production in photon-nucleus
interactions based on the {\sc Phojet}-model for the description of 
photon-nucleon interactions. There, we will argue that a model which should
give a reasonable description of photoproduction off nuclei has to be able
to describe the main features of photon-proton interactions as well.
Since it was found by the H1-Collaboration~\cite{Aid95c} that the 
{\sc Phojet}
event generator provides a reasonable description of $\gamma p$ 
photoproduction at 200 GeV c.m. energy using a transverse momentum cutoff of 
3~GeV/$c$ we use this value consistently for the calculation of cross 
sections
and of particle production in photon-proton and in photon-nucleus 
interactions.
The formalism presented here is only applicable for photons with low $Q^2$,
i.e. $Q^2\ll 4p_\perp^2$.
In Fig.~\ref{gptot} the calculated cross sections for direct and anomalous 
photon interactions on a proton target are shown.
\section{Photon-nucleus interactions}
\subsection{\label{xs_gA}Cross sections}
The application of Eq.(\ref{GVDM-sig_gn}) to the scattering of a
virtual photon on a nuclear target of mass number
$A$ is straightforward (see~\cite{Piller95} and references therein). 
In order to calculate the total virtual photon-nucleus cross section
$\sigma_{\gamma^{\star}A}$, $\sigma_{VN}$ has to be replaced by the
effective cross section $\sigma_{VA}$ for the interaction of a 
$q\bar{q}$-system of mass $M$ with a nucleus with mass number $A$:
\begin{equation}
\label{GVDM-sig_gA}
\sigma_{\gamma^{\star}A}(s,Q^2)=4\pi\alpha_{\mbox{\scriptsize em}} 
\int_{M_0^2}^{M_1^2} dM^2\ D(M^2) \left(\frac{M^2}{M^2+Q^2}\right)^2
\left(1+\epsilon\frac{Q^2}{M^2}\right) \sigma_{VA}(s,Q^2,M^2).
\end{equation}

$\sigma_{VA}$ is obtained as follows:
For coherence lengths $d$ of the hadronic fluctuation
\begin{equation}
\label{d_coh}
d=\frac{2\nu}{M^2+Q^2}
\end{equation}
exceeding the average distance between two nucleons the $q\bar{q}$-system
may interact coherently with several nucleons of the target nucleus. 
This multiple scattering process can be described using the MC realization
of the Glauber-Gribov approximation by 
Shmakov {\it et al.}~\cite{Shmakov89}, here, extended to photon
projectiles. The high energy small-angle scattering amplitude $F$ for the 
interaction of
a $q\bar{q}$-system with a nucleus at impact parameter $\vec{b}$
can be written in terms of the impact parameter amplitude $\Gamma$ for the
interaction of the $q\bar{q}$-system with individual 
nucleons~\cite{Shmakov89}
\begin{equation}
\label{Gamma_gl}
F(\vec{b})=\langle \psi_A^f|1-\prod_{i=1}^{A}\left[1-
\Gamma(\vec{b}_i)\right]|\psi_A^i\rangle, \qquad
\vec{b}_i=\vec{b}-\vec{s}_i.
\end{equation}
The $\vec{s}_i$ are the coordinates of the nucleons with regard to the
center of mass of the nucleus in the plane of impact parameter. The
scattering amplitude is averaged over the initial and final state wave
functions $\psi_A^i$ and $\psi_A^f$ of the nucleus. For the
$q\bar{q}$-nucleon scattering amplitude we assume the following
parametrization
\begin{equation}
\label{gamma}
\Gamma(s,Q^2,M^2,\vec{b})=
\frac{\sigma_{VN}(s,Q^2,M^2)}{4\pi B(s,Q^2,M^2)}
\left(1-i\frac{\Re e f(0)}{\Im m f(0)}\right) 
\exp{\left(\frac{\vec{b}^2}{2B(s,Q^2,M^2)}\right)}.
\end{equation}
The effective $q\bar{q}$-nucleon cross section $\sigma_{VN}$ is obtained as 
discussed in Sect.\ref{Sect_sig-gn} (Eq.(\ref{GVDM-sig_gn-sig_VN})).
We adopt the parametrization of the slope $B$ from~\cite{Haakman96}
\begin{eqnarray}
\label{slope_a}
B(s,Q^2,M^2)=2 \left[B_0^2+\alpha_{I\!\!P}'\ln \left(\frac{s}{M^2+Q^2}\right)
\right], \nonumber \\
B_0^2=\left( 2+\frac{m_{\rho}^2}{M^2+Q^2}\right)  \ \ {\mbox{\rm GeV}}^{-2},
\qquad \alpha_{I\!\!P}'=0.25\ \ {\mbox{\rm GeV}}^{-2},
\end{eqnarray}
and assume for the ratio $\Re e f(0)/\Im m f(0)$ a constant value of 0.1.
Neglecting correlations between nucleons one may write
\begin{equation}
\label{rho_A}
|\psi_A^i|^2=\prod_{j=1}^{A}\rho_A(\vec{s}_j,z_j),\qquad
\rho_A(\vec{r})=\frac{K}{1+\exp{\left[(|\vec{r}|-R_A)/c\right]}}
\end{equation}
where $\rho_A$ is the one-particle Woods-Saxon density distribution
with $c=0.545$~fm and $R_A=1.12 A^{1/3}$~fm~\cite{Segre77}. 
Therefore, for the total, inelastic, and elastic $q\bar{q}$-nucleus cross 
sections $\sigma_{VA}^{\mbox{\scriptsize tot}}$,
$\sigma_{VA}^{\mbox{\scriptsize inel}}$, and 
$\sigma_{VA}^{\mbox{\scriptsize el}}$ we obtain
\begin{eqnarray}
\label{GVDM-sig_VAtot}
\sigma_{VA}^{\mbox{\scriptsize tot}}(s,Q^2,M^2) &=& 
2 \Re e \int d^2b\ F(s,Q^2,M^2,\vec{b}) \nonumber \\
&=& 2 \Re e \left\{ \int d^2b\ \int \prod_{j=1}^{A} d^3r_j\
\rho_A(\vec{r}_j) \left(1-\prod_{i=1}^{A}\left[1-\Gamma(s,Q^2,M^2,\vec{b}_i)
\right] \right) \right\}, \\
\label{GVDM-sig_VAinel}
\sigma_{VA}^{\mbox{\scriptsize inel}}(s,Q^2,M^2) &=& 
\int d^2b\ \left(1-\left|1-F(s,Q^2,M^2,\vec{b})\right|^2\right) 
\nonumber \\
&=& \int d^2b\ \int \prod_{j=1}^{A} d^3r_j\
\rho_A(\vec{r}_j) \left(1-
\left|\prod_{i=1}^{A}\left[1-\Gamma(s,Q^2,M^2,\vec{b}_i)
\right]\right|^2 \right), \\
\label{GVDM-sig_VAel}
\sigma_{VA}^{\mbox{\scriptsize el}}(s,Q^2,M^2) &=& 
\sigma_{VA}^{\mbox{\scriptsize tot}}(s,Q^2,M^2)-
\sigma_{VA}^{\mbox{\scriptsize inel}}(s,Q^2,M^2).
\end{eqnarray}
The integrations over the $\vec{r}_j$'s are performed by taking the average 
of the integrand in Eqs.(\ref{GVDM-sig_VAtot},\ref{GVDM-sig_VAinel}) over 
a sufficiently large number of nucleon coordinates sets
sampled from the density distribution $\rho_A$.

At low energies the coherence length $d$ may lead to a suppression
of shadowing which we take into account in the calculation of the product 
over the $A$ nucleons for a fixed spatial nucleon configuration
(Eqs.(\ref{GVDM-sig_VAtot},\ref{GVDM-sig_VAinel})). 
In the non-shadowing limit, i.e. if $d$ is smaller than any internucleon
distance, we obtain a sum over $\Gamma(\vec{b}_i)$ and, therefore, 
$\sigma_{VA}\approx A \sigma_{VN}$.

As discussed initially, in direct photon interactions the Glauber multiple
scattering process is, per definition, completely suppressed since the 
photon couples directly to a parton in a nucleon without leaving any remnant.
In contrast, the assumption that also in anomalous photon interactions the
Glauber cascade is reduced to one $q\bar q$-nucleon scattering can only be 
considered as an extreme case. However, this might be justified since we are 
interested in estimating the maximum possible effect of the point-like
interactions on the shadowing behaviour at high energies. 
We recognize however, that the soft component of hard processes has been
discussed (see for instance~\cite{Kopeliovich96b}).
It is obvious,
how the point-like processes have to be taken into account in 
Eq.(\ref{gamma}):
$\sigma_{VN}$ has to be replaced by $(1-\xi)\sigma_{VN}$ and 
$A\cdot\sigma^{\mbox{\scriptsize pl}}_{\gamma^{\star}N}$ is added explicitly
to Eq.(\ref{GVDM-sig_gA}), with
\begin{equation}
\label{fracpl}
\xi(s,Q^2)=\frac{\sigma_{\gamma^{\star}N}^{\mbox{\scriptsize pl}}(s,Q^2)}
               {\sigma_{\gamma^{\star}N}^{\mbox{\scriptsize tot}}(s,Q^2)},
\qquad
\sigma^{\mbox{\scriptsize pl}}_{\gamma^{\star}N}(s,Q^2) =
\sigma^{\mbox{\scriptsize dir}}_{\gamma^\star N}(s,Q^2)+
\sigma^{\mbox{\scriptsize ano}}_{\gamma^\star N}(s,Q^2).
\end{equation}

In Fig.~\ref{gAtot} we compare our results on $\sigma_{\gamma A}$ in the
photoproduction limit ($Q^2=0$) for carbon, copper, and lead
targets (solid lines) to 
data~\cite{Arakelyan78,Brookes73,Caldwell73,Caldwell79}. The
agreement is reasonable apart from the low energy region where our
results for the carbon target seem to show less shadowing than measured.
However, for $\nu\approx 2-3$~GeV the lower energy limit of the applicability
of the model is reached.
In addition, we indicate with dotted lines the cross sections which would 
be obtained if one neglects the limited coherence length at low energies.
Whereas this effect is less significant for light targets, it is
responsible for the increase of the cross sections towards lower
energies observed in interactions of real photons with copper and lead 
nuclei.

An extrapolation in energy of the real photon-nucleus cross section is
presented in Fig.~\ref{gAtotcms}, again, for the three target nuclei
carbon, copper, and lead. Here, the point-like interactions lead
to a stronger increase of the cross sections above a photon-nucleon
c.m. energy of 100~GeV (solid lines) than it would be obtained neglecting
the suppression of the Glauber-cascade by point-like processes (dotted
lines). 
\subsection{Nuclear shadowing of photons}
The ratio of the total photon-nucleus to the total photon-nucleon cross
section, which gives the effective number of nucleons $A_{\mbox{\scriptsize
eff}}$ ``seen'' by the photon projectile, has been measured in 
photoproduction experiments using carbon, copper, and lead 
targets~\cite{Arakelyan78,Brookes73,Caldwell73,Caldwell79}.
We compare these data in the form $A_{\mbox{\scriptsize eff}}/A$, frequently
called ``effective attenuation'', to results of our calculations in 
Fig.~\ref{gAsha}a-c. The agreement is reasonable. However, there are
considerable uncertainties within the measurements as well as differences
between the results obtained in different experiments which make it 
difficult to draw further conclusions from this comparison.

In Fig.~\ref{gAshacms} the shadowing ratios 
$\sigma_{\gamma^{\star}A}/(A\sigma_{\gamma^{\star}N})$
for real photons are extrapolated 
in energy up to a photon-nucleon c.m. energy of 2~TeV. In order to study
the influence of point-like processes to the high energy shadowing behaviour
we plot the full model (solid lines) and the cross sections obtained if
the point-like processes are not taken into consideration (dotted lines).
From this comparison we conclude that point-like processes are
responsible for a decrease of the nuclear shadowing  with increasing energy.

Let us now turn to lepton-nucleus interactions where the shadowing 
region ($x<0.1$) has been investigated by the E665-Collaboration using
470~GeV/$c$ muons and by the NMC-Collaboration using 200~GeV
muons. Within our calculations the flux $g$ of
virtual photons is sampled according to the equivalent photon
approximation (EPA) folded with the $Q^2$-dependent cross section
$\sigma_{\gamma^{\star}A}$ (Eq.(\ref{GVDM-sig_gA}), see 
also~\cite{Engel95d} for details)
\begin{equation}
\sigma_{lA}=\int dy\ \int dQ^2\ g(y,Q^2) \sigma_{\gamma^{\star}A}(s,Q^2),
\end{equation}
with $y$ being the lepton energy fraction taken by the photon.
The kinematic cuts as they were applied to the measured data are taken
into account.
In Fig.~\ref{gvAsha_E665}a-c we compare the model predictions concerning
the $x$-dependence of the cross section ratios to E665-~\cite{Adams95} and 
NMC data~\cite{Amaudruz91,Arneodo95}. Our results are binned in the
same way as the E665 data showing that our photon-flux approximation 
gives average $x$-values in each bin which correspond to the 
measured ones. The average $Q^2$-values range from 0.15~GeV$^2$ in the
lowest $x$-bin up to 7.9~GeV$^2$ in the highest bin. The calculations
for the three target nuclei carbon, calcium, and lead are in good
agreement with the E665 data but overestimate the NMC data slightly.

In order to study the $Q^2$-dependence of the cross section ratios
at fixed values of $x$ we parametrize them as
\begin{equation}
R^A=\frac{\sigma_{\gamma^{\star}A}}{A\sigma_{\gamma^{\star}N}}
=a+b\cdot\log(Q^2/{\mbox{\rm GeV}}^2).
\end{equation}
In Fig.~\ref{gvAsha_b_E665} we plot the slope $b$ as
function of $x$, again for carbon, calcium, and lead targets, together with 
E665-measurements~\cite{Adams95}.
Our results are consistent with the experimental observations, i.e. with 
a weak $Q^2$-dependence of the shadowing effect within the considered 
$x$-range.

The strength of shadowing may also be studied by parametrizing the 
per-nucleon cross section
ratios by $R^A\propto A^{\alpha-1}$.
In Fig.~\ref{gvAalpha_E665} we compare results of our calculations on the 
values of $\alpha$ to E665 data~\cite{Adams95}. Even though our values are
systematically above the data, they are still compatible with them.

Extrapolating the shadowing ratios to high energies at fixed large values 
of $Q^2$ ($x\to 0$) Kopeliovich and Povh predicted within their model 
that shadowing vanishes~\cite{Kopeliovich96a}.
As shown in Fig.~\ref{gvAsha_kop}
for carbon (a) and lead targets (b), within our model we predict
the same qualitative features since the soft contributions 
are stronger suppressed with the photon virtuality than the point-like 
contributions. Applying an energy-independent cutoff to calculate the 
point-like photon interactions, one would get also a decrease of shadowing in 
the photoproduction limit at very high energies. However, it is expected that
the transverse momentum cutoff should increase with energy in order to
guarantee that the calculation is restricted to a kinematic region where
lowest-order perturbative QCD estimates are reliable. Several 
parametrizations of the energy-dependence of the cutoff have been suggested
in~\cite{Geiger93,Bopp94a}. These parametrizations predict a only 
slowly varying cutoff up to c.m. energies of about 2~TeV. Therefore, 
we assume
that the qualitative results reported here do not change applying an 
energy-dependent cutoff. This has been confirmed numerically for the 
parametrization discussed in~\cite{Bopp94a}.
\subsection{Quasi-elastic vector meson production}
A further test of the $Q^2$-behaviour of our model can be performed by
studying quasi-elastic vector meson production. For example, the cross
section for the (coherent) quasi-elastic $\rho^0$ production off a nucleus 
with mass number $A$ depends on $Q^2$ like
\begin{equation}
\sigma_{\gamma^{\star}A\longrightarrow\rho^0 A}(s,Q^2)\sim
\left(\frac{m_\rho^2}{m_\rho^2+Q^2}\right)^2
\left(1+\epsilon\frac{Q^2}{m_\rho^2}\right)
\sigma_{\rho^0A}^{\mbox{\scriptsize el}}(s,Q^2)
\end{equation}
where $\sigma_{\rho^0A}^{\mbox{\scriptsize el}}$ is obtained according to
(\ref{GVDM-sig_VAel}).
In Fig.~\ref{gAqelrho} we show our results together with data of the
NMC-Collab.~\cite{Arneodo94a}. In each $Q^2$-bin, the average photon
energy and the average value of $\epsilon$ were used as given 
in~\cite{Arneodo94a}. In order to compare the shape, our results were 
normalized to the data. Parametrizing 
$\sigma_{\rho^0A}^{\mbox{\scriptsize el}}$ as
\begin{equation}
\sigma_{\rho^0A}^{\mbox{\scriptsize el}}(Q^2)=\sigma_0\left(
\frac{Q_0^2}{Q^2}\right)^\beta
\end{equation}
we get from a fit to our results values for $\beta$ of $2.6$ for
deuterium, $2.5$ for carbon, and $2.4$ for calcium. 
Since our results slightly deviate from a power-law behaviour in $Q^2$
we estimate the uncertainties for the $\beta$-values by excluding either
the cross sections at the two lowest or highest $Q^2$ from the fit.
For the three $\beta$-values we obtain an uncertainty of $\pm 0.2$.
Comparing it to the experimental value of $\beta=2.02\pm 0.07$ from a
combined fit to the data for the three target nuclei~\cite{Arneodo94a},
we conclude that our results show a somewhat stronger $Q^2$-dependence than 
the NMC data.
%
%
\section{Summary and conclusions}
Cross sections for photon-nucleus interactions are calculated based on the
assumption that the photon may interact directly or as a ``resolved'' 
$q\bar{q}$-state. We apply the Generalized Vector Dominance Model together 
with Glauber-Gribov theory taking coherence length effects into account. 
Total photon-nucleon cross sections are calculated within the CKMT-model.

We assume that direct and anomalous photon interactions are point-like, 
i.e. the photon interacts with only one target nucleon. The cross sections 
corresponding to the point-like processes are estimated within lowest order 
perturbative QCD. The suppression of the Glauber cascade due to these
point-like photon interactions is explicitly taken into account.

Real photon-nucleus cross sections are compared to data and extrapolated
to high energies. In addition, we study the shadowing behaviour in 
photon-nucleus interactions for moderate photon virtualities by comparing
them to data. In both cases a good agreement with the data is found.
It is discussed that point-like photon interactions lead to a
suppressed shadowing in interactions with heavy target nuclei at high 
energies.

In a forthcoming paper we will extend this study to particle production in
interactions of weakly virtual photons off nuclei. There, the description
of particle production will be based on the above discussed cross sections
and shadowing behaviour combined with the ideas of the two-component Dual 
Parton Model describing photon-nucleon interactions.
%
%
\section*{Acknowledgements}
Discussions with F.\ W. Bopp are gratefully acknowledged.
One of the authors (J.R.) thanks C.\ Pajares for the hospitality
at the University Santiago de Compostela and he was supported by the
Direccion General de Politicia Cientifica of Spain.
One of the authors (R.E.) was supported by the Deutsche 
Forschungsgemeinschaft under contract No. Schi 422/1-2.
%
%

%
%
\clearpage
\section*{Figure Captions}
\begin{enumerate}
\item \label{gptot} 
      Total photon-proton cross sections as calculated with the
      CKMT-model~\protect\cite{Capella94b} (thick solid line) and
      obtained within the two-component DPM (dotted line) are shown together 
      with measurements~\protect\cite{Alekhin87,Derrick94b,Aid95b}. 
      In addition, we give the contribution to the
      total cross section from direct processes and the cross section
      reflecting the anomalous component of the photon-PDFs.
\item \label{gpsigvn}
      The effective $q\bar{q}$-nucleon cross sections at $M^2=m_{\rho}^2$
      are shown. In (a) the dependence on the energy is given for three
      different photon virtualities. In (b) we show the $Q^2$-behaviour
      for three different energies.
\item \label{grf_plike}
      Examples for point-like interactions of the photon. In (a) direct
      photon-nucleon interactions contributing in lowest order $p$QCD
      and in (b) an example for an anomalous photon-nucleon interaction 
      are shown.
\item \label{gAtot} 
      The dependence of the total cross sections for interactions of real 
      photons with carbon, copper, and lead on the photon energy (full lines)
      is compared to 
     measurements~\protect\cite{Arakelyan78,Brookes73,Caldwell73,Caldwell79}.
      The influence of the coherence length is indicated by 
      dotted lines where we show the cross sections as they would be 
      obtained disregarding the finite coherence length of the photon.
\item \label{gAtotcms} 
      Extrapolation of total cross sections for photoproduction off
      carbon-, copper-, and lead-nuclei. $\sqrt{s}$ is the
      photon-nucleon c.m. energy. The pure GVDM-prediction is shown by
      the dotted lines. The cross sections taking the suppression 
      of shadowing by point-like photon-nucleon interactions 
      into consideration are given by solid lines.
\item \label{gAsha} 
      Per-nucleon ratios of real photon-carbon (a), -copper (b), and -lead 
      (c) cross sections to photon-nucleon cross sections are shown
      together with 
      measurements~\protect\cite{Arakelyan78,Brookes73,Caldwell73,Caldwell79}.
\item \label{gAshacms} 
      As in Fig.~\ref{gAtotcms} but for the shadowing ratios.
\item \label{gvAsha_E665} 
      The dependence of the per-nucleon ratios of photon-carbon (a), 
      -calcium (b), and -lead (c) cross sections to photon-nucleon cross 
      sections on the Bjorken-$x$ is compared to data of the
      E665-~\protect\cite{Adams95} and 
      NMC-Collab.~\protect\cite{Amaudruz91,Arneodo95}.
\item \label{gvAsha_b_E665} 
      The slopes of the logarithmic $Q^2$-dependence of the shadowing
      ratio $R^A$ are
      compared to results of the E665-Collab.~\protect\cite{Adams95}.
\item \label{gvAalpha_E665} 
      The $A$-dependence of the per-nucleon cross section ratios
      $R^A\propto A^{\alpha-1}$ for different
      bins of the Bjorken-$x$ is shown together with
      E665 data~\protect\cite{Adams95}.
\item \label{gvAsha_kop} 
      Bjorken-$x$-dependence of the per-nucleon ratios of photon-carbon (a)
      and photon-lead (b) cross sections to photon-nucleon cross sections 
      shown for $Q^2=0.1, 0.5, 1.0$, and $5.0$~GeV$^2$ (from the bottom
      to the top).
\item \label{gAqelrho} 
      Dependence of the quasi-elastic $\rho^0$-production cross sections
      on the photon virtuality for photon-deuterium (a), -carbon (b),
      and -calcium (c) interactions. The model results are normalized to
      the data of the NMC-Collab.~\cite{Arneodo94a}.
\end{enumerate}
\clearpage
\newpage
\pagestyle{empty}
\begin{figure}[htb]
\setlength{\unitlength}{1cm}
\begin{picture}(15,23)(0,0)
\put(-2.0,0.0){\psfig{figure=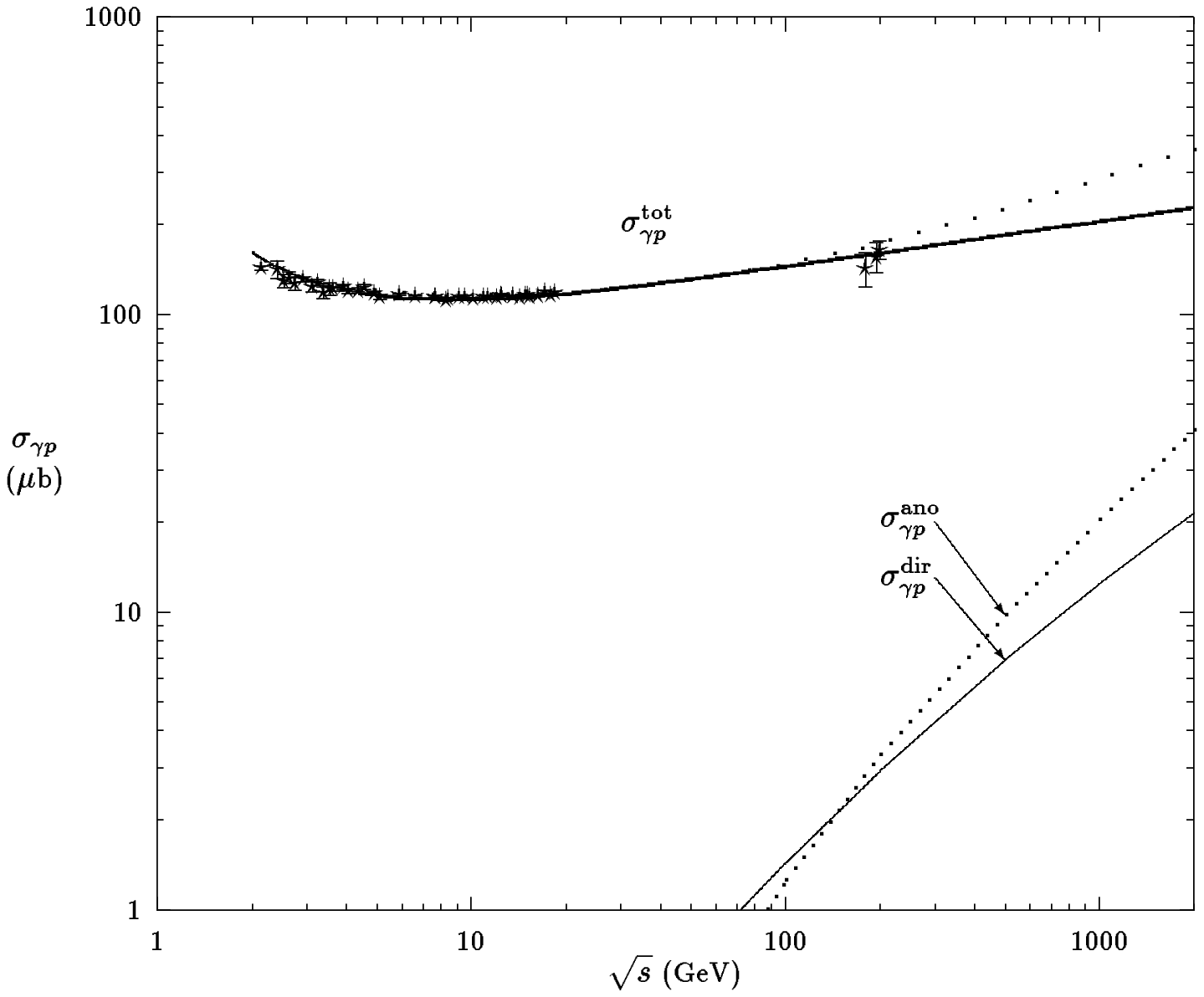}}
\put(7.5,0.0){\bf\large Fig.~\ref{gptot}}
\end{picture}
\end{figure}
\clearpage
\newpage
\begin{figure}[htb]
\setlength{\unitlength}{1cm}
\begin{picture}(15,23)(0,0)
\put(-2.0,7.5){\psfig{figure=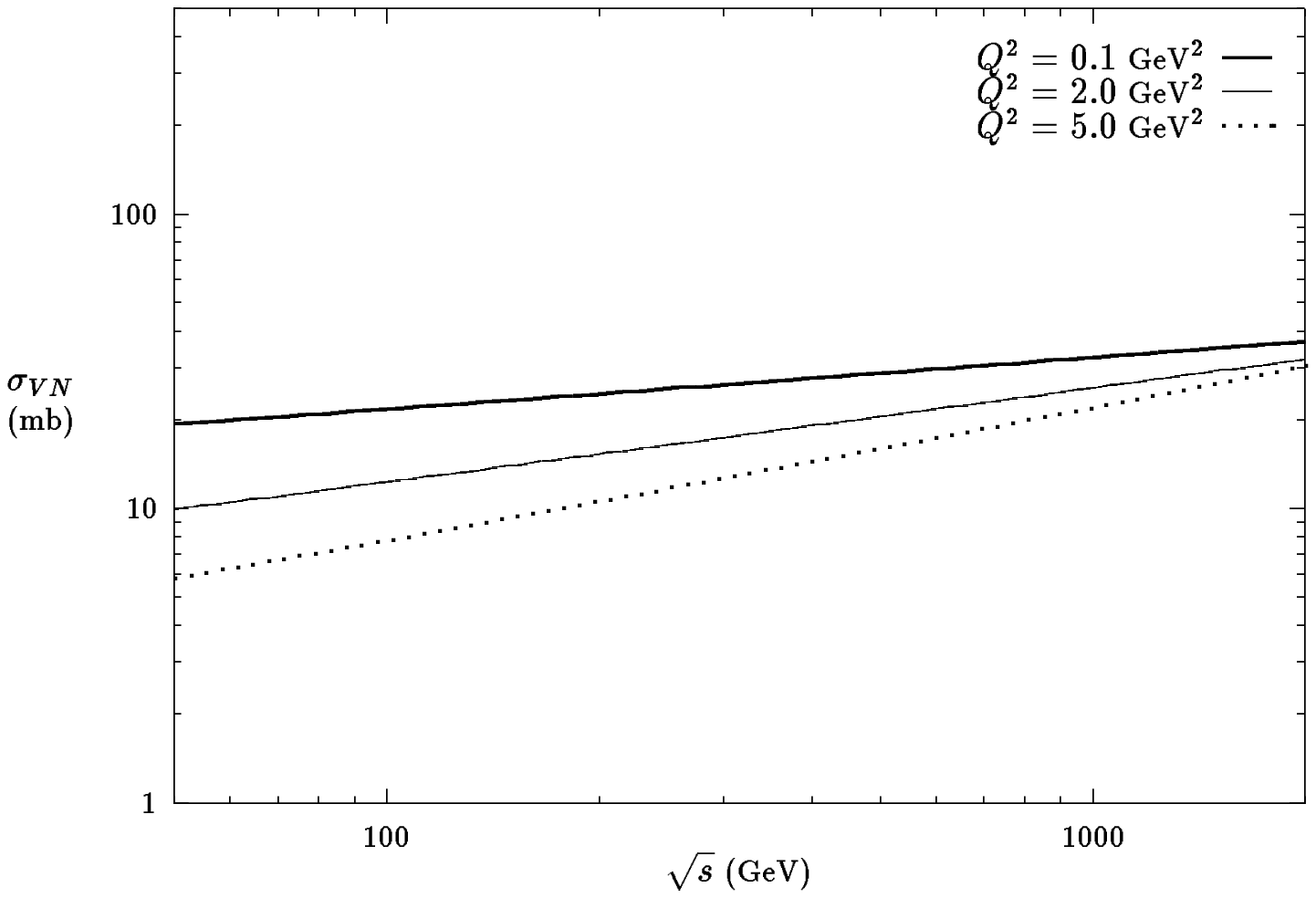}}
\put(0,13.0){\bf\large a)}
\put(-2.0,-4.5){\psfig{figure=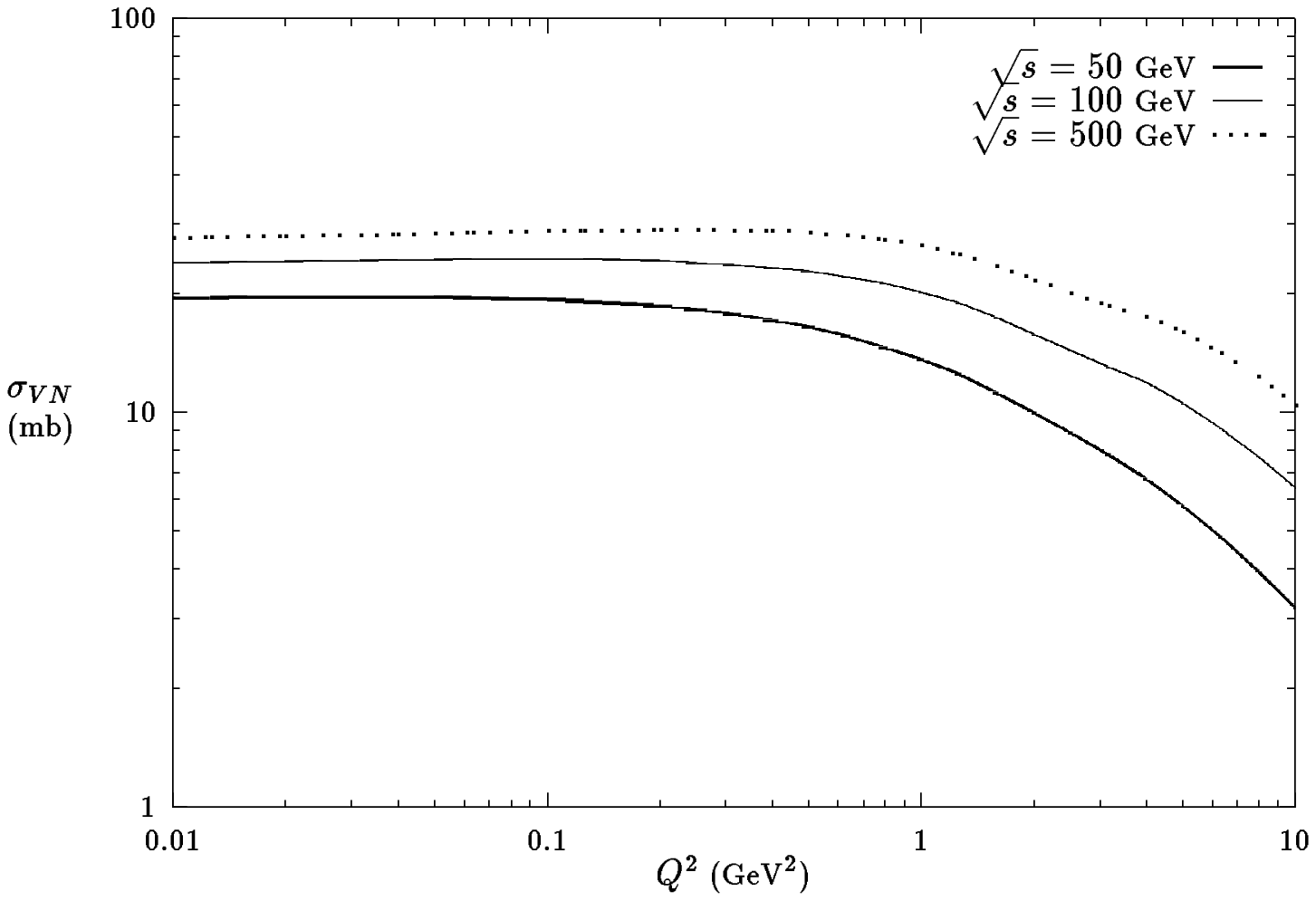}}
\put(0,1.0){\bf\large b)}
\put(7.5,0.0){\bf\large Fig.~\ref{gpsigvn}}
\end{picture}
\end{figure}
\clearpage
\newpage
\begin{figure}[htb]
\setlength{\unitlength}{1cm}
\begin{picture}(15,23)(0,0)
\put(2.5,13.5){\psfig{figure=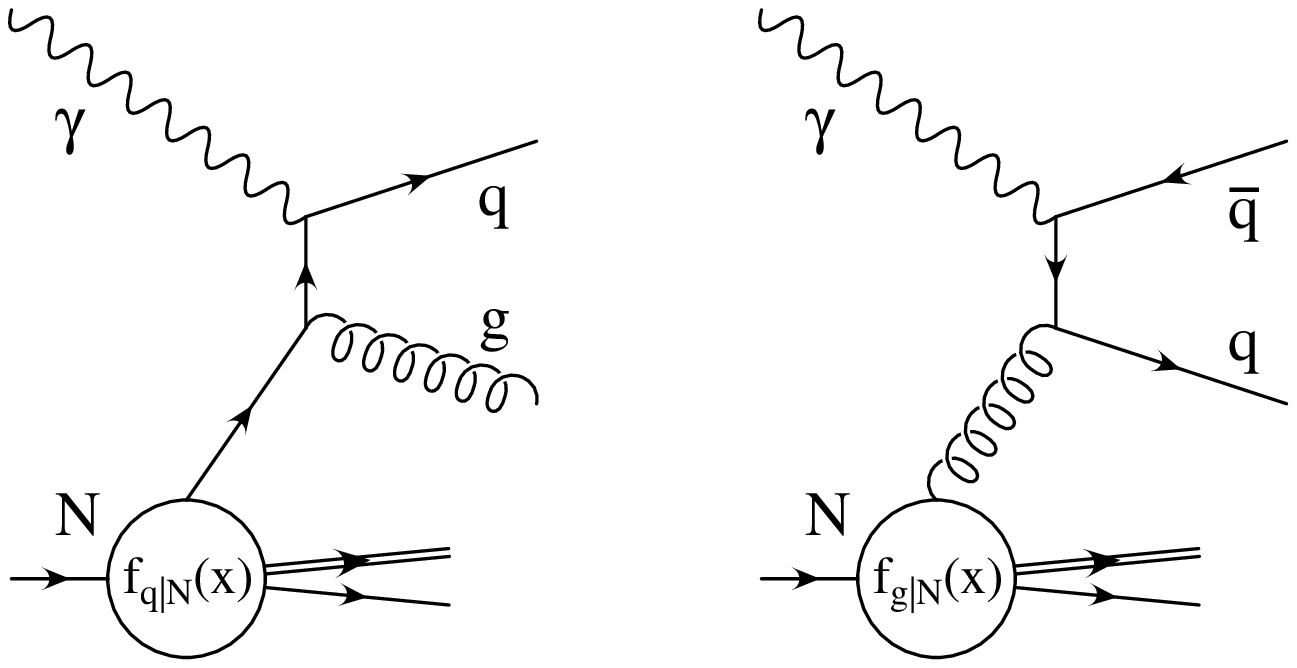}}
\put(3,12.0){\bf\large a)}
\put(6.5,4.5){\psfig{figure=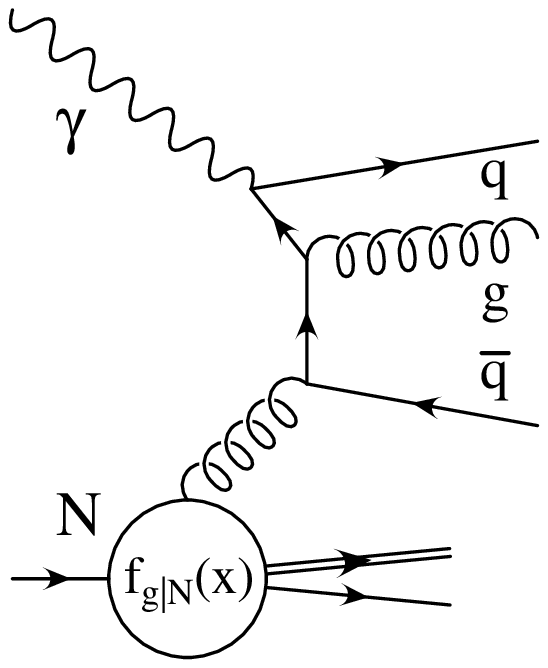}}
\put(3,3.0){\bf\large b)}
\put(7.5,0.0){\bf\large Fig.~\ref{grf_plike}}
\end{picture}
\end{figure}
\clearpage
\newpage
\begin{figure}[htb]
\setlength{\unitlength}{1cm}
\begin{picture}(15,23)(0,0)
\put(-2.0,0.0){\psfig{figure=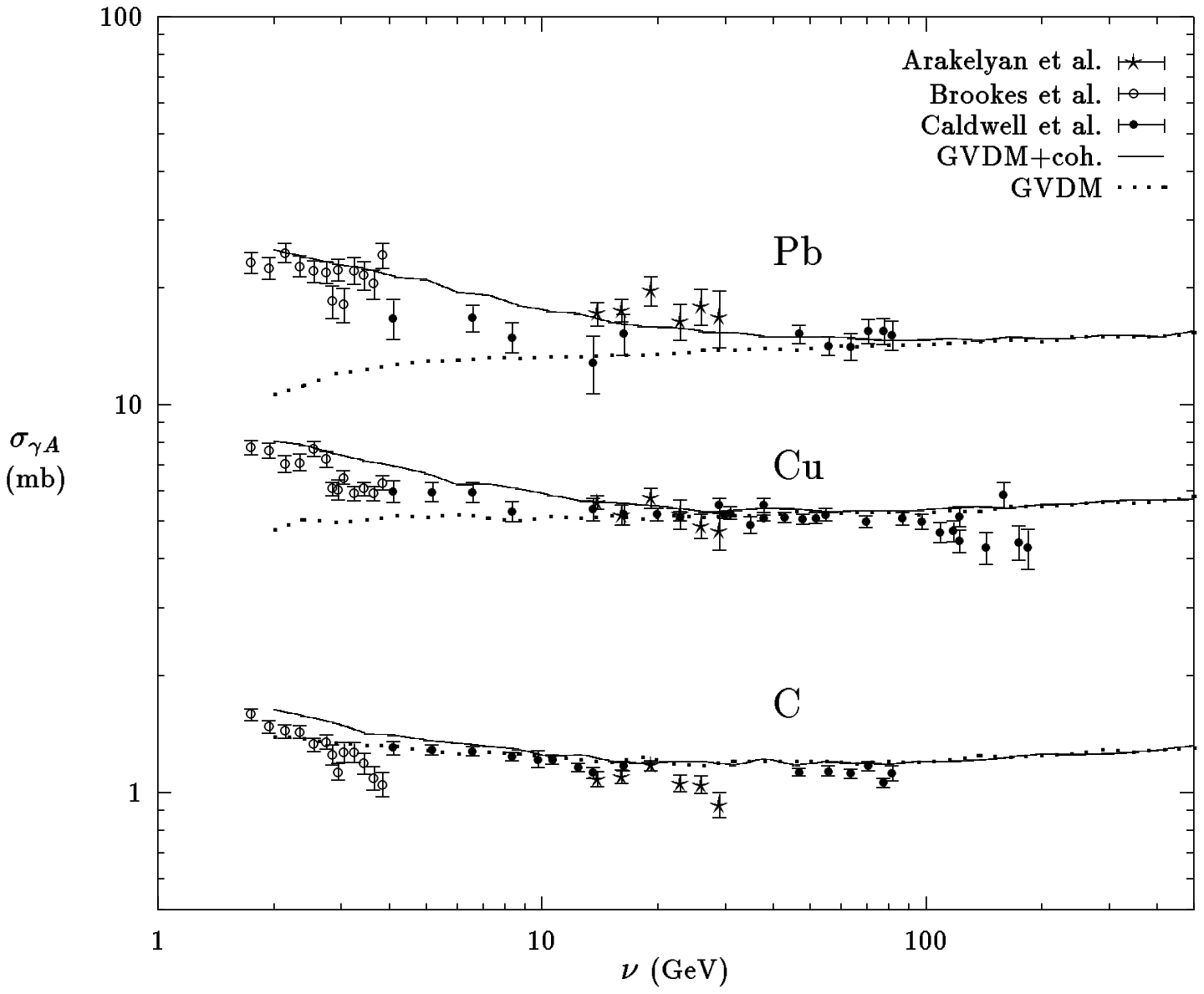}}
\put(7.5,0.0){\bf\large Fig.~\ref{gAtot}}
\end{picture}
\end{figure}
\clearpage
\newpage
\begin{figure}[htb]
\setlength{\unitlength}{1cm}
\begin{picture}(15,23)(0,0)
\put(-2.0,0.0){\psfig{figure=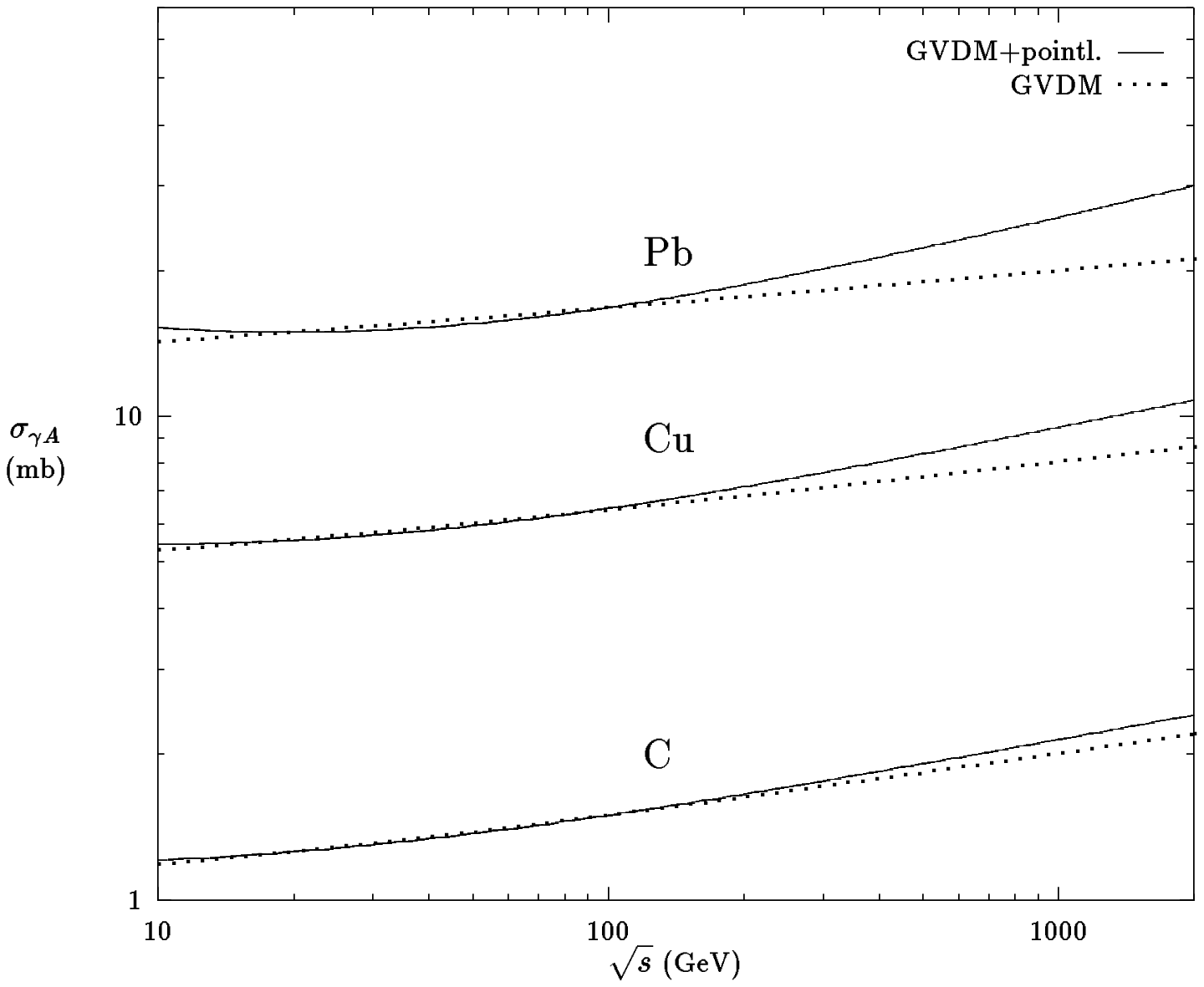}}
\put(7.5,0.0){\bf\large Fig.~\ref{gAtotcms}}
\end{picture}
\end{figure}
\clearpage
\newpage
\begin{figure}[htb]
\setlength{\unitlength}{1cm}
\begin{picture}(15,23)(0,0)
\put(-0.5,10.0){\psfig{figure=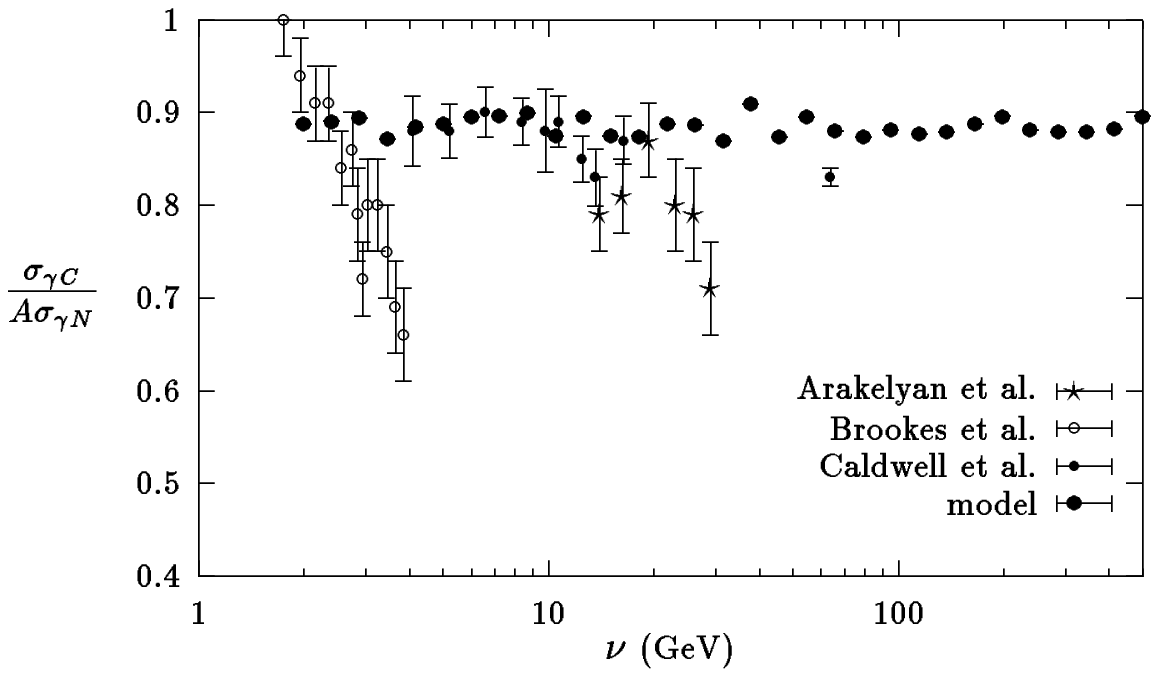}}
\put(3.0,16.0){\bf\large a)}
\put(-0.5,2.5){\psfig{figure=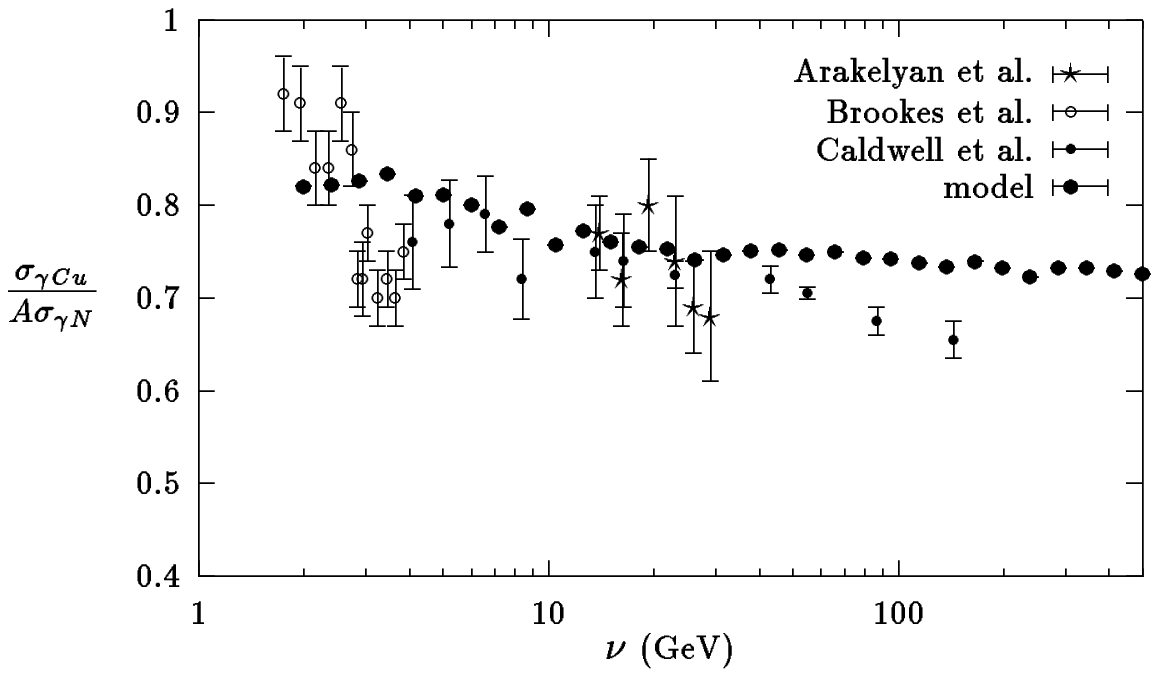}}
\put(3.0,8.5){\bf\large b)}
\put(-0.5,-5.0){\psfig{figure=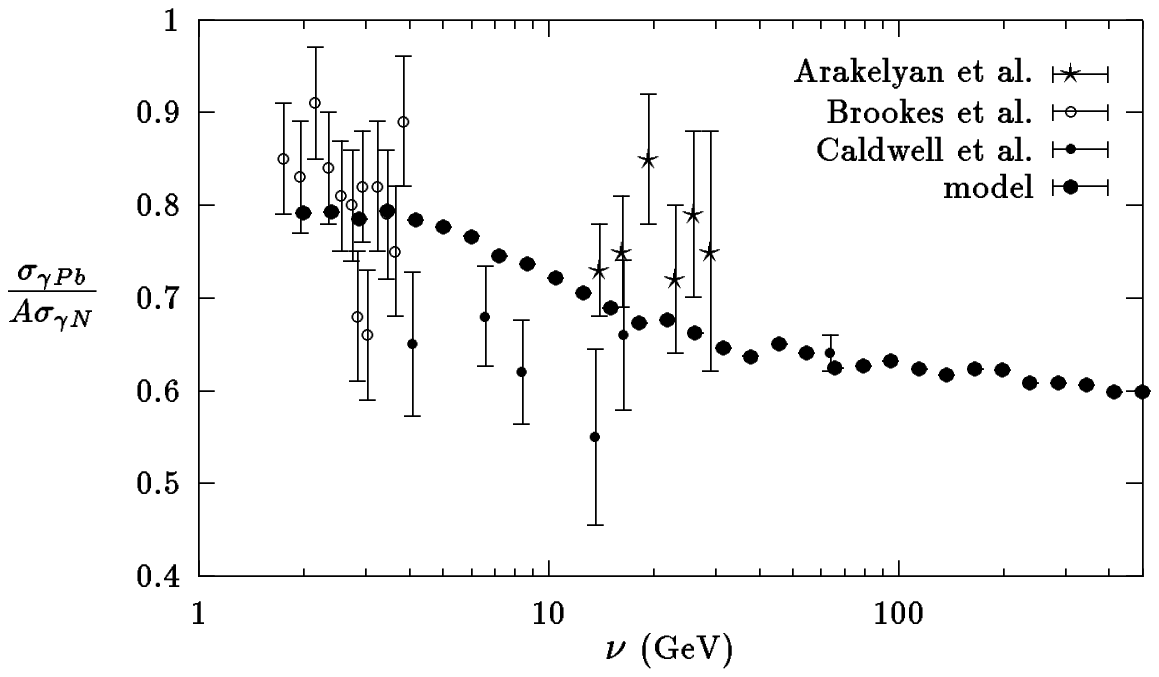}}
\put(3.0,1.0){\bf\large c)}
\put(7.5,0.0){\bf\large Fig.~\ref{gAsha}}
\end{picture}
\end{figure}
\clearpage
\newpage
\begin{figure}[htb]
\setlength{\unitlength}{1cm}
\begin{picture}(15,23)(0,0)
\put(-2.0,0.0){\psfig{figure=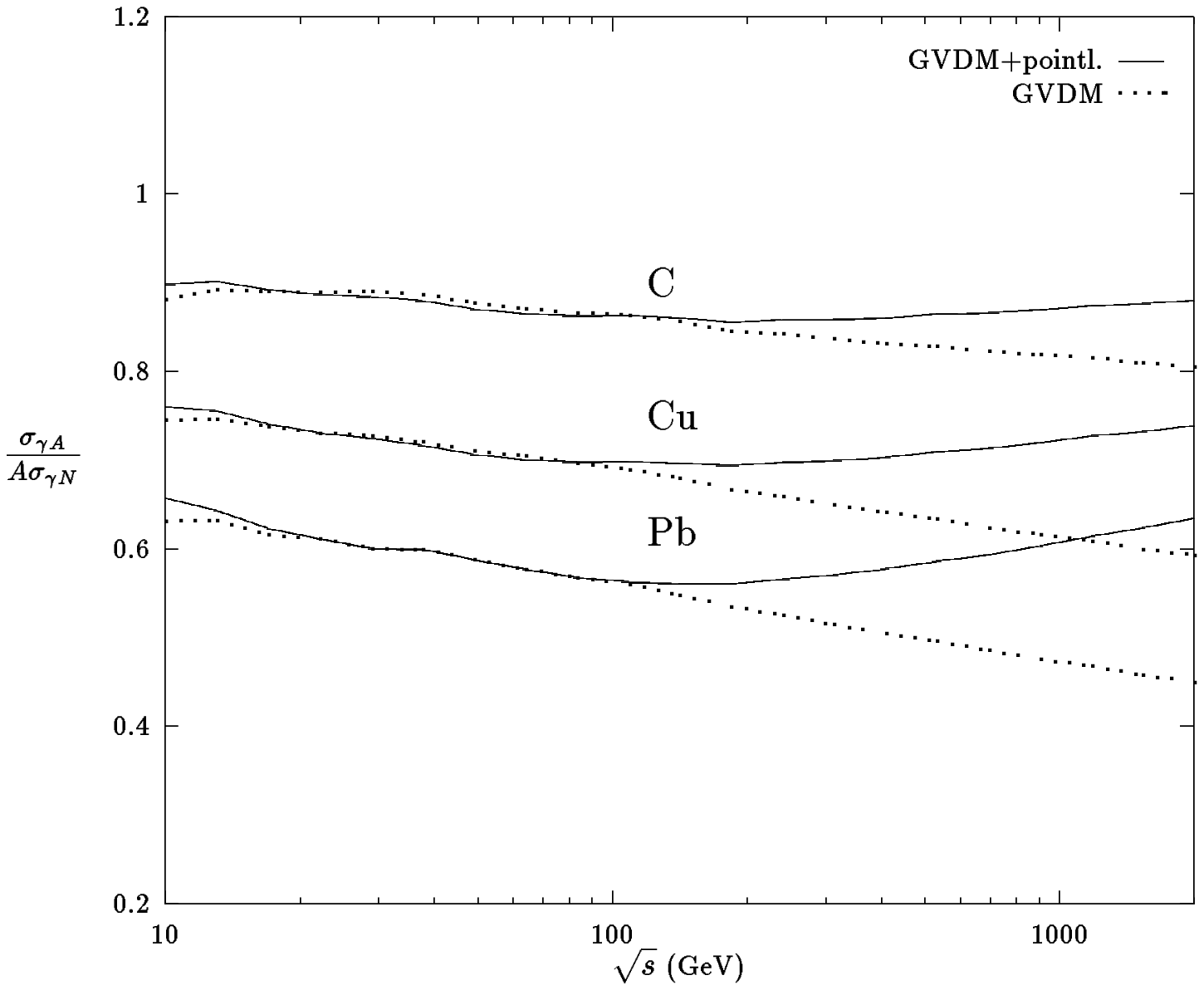}}
\put(7.5,0.0){\bf\large Fig.~\ref{gAshacms}}
\end{picture}
\end{figure}
\clearpage
\newpage
\begin{figure}[htb]
\setlength{\unitlength}{1cm}
\begin{picture}(15,23)(0,0)
\put(-0.5,10.0){\psfig{figure=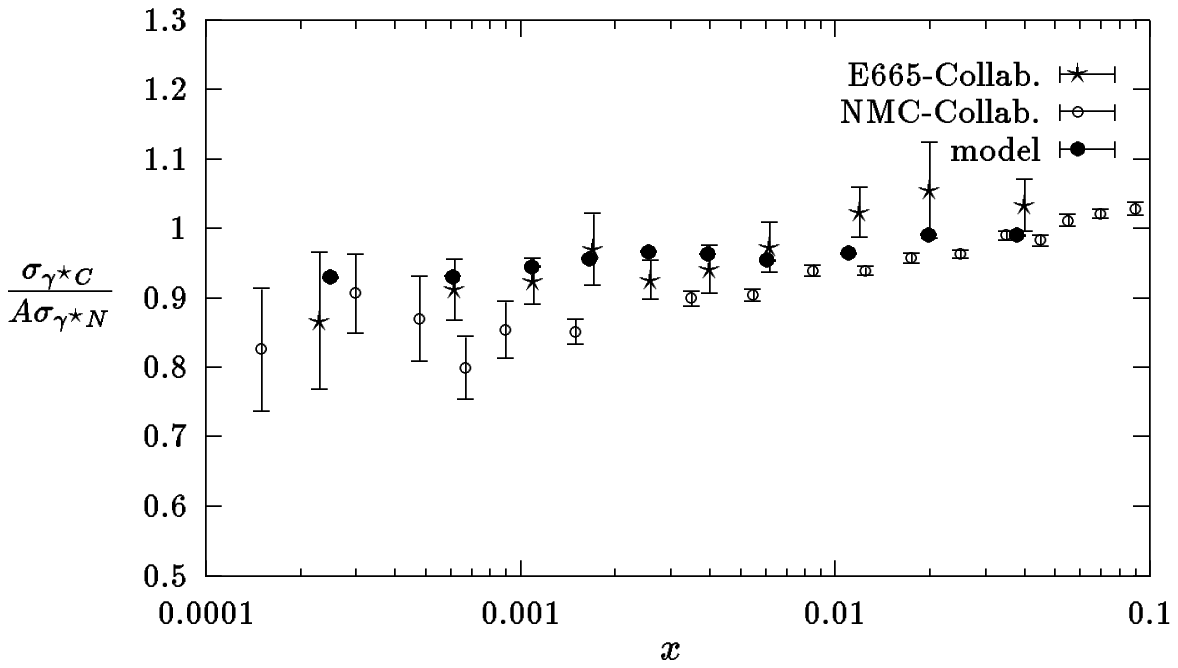}}
\put(3.0,16.0){\bf\large a)}
\put(-0.5,2.5){\psfig{figure=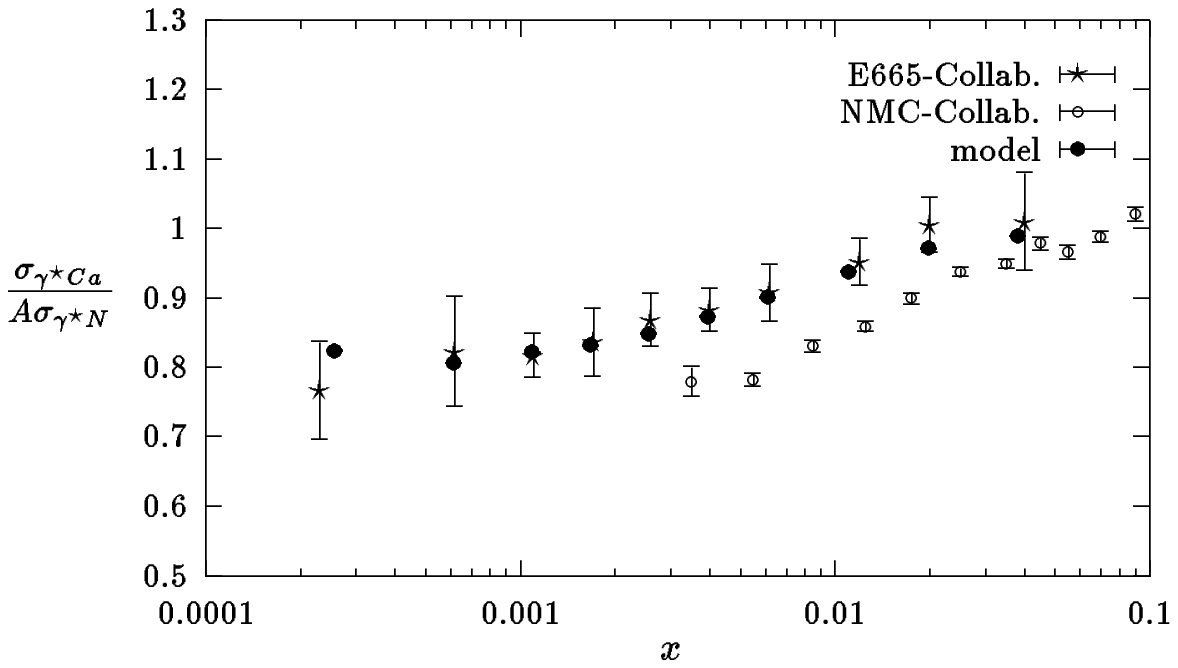}}
\put(3.0,8.5){\bf\large b)}
\put(-0.5,-5.0){\psfig{figure=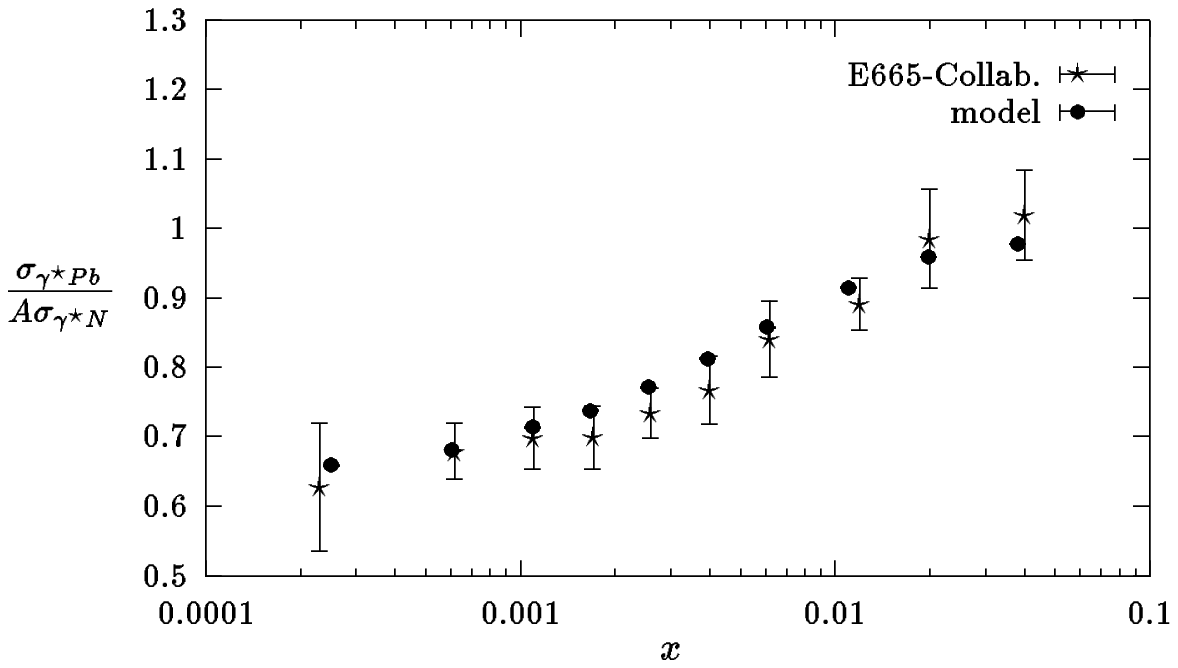}}
\put(3.0,1.0){\bf\large c)}
\put(7.5,0.0){\bf\large Fig.~\ref{gvAsha_E665}}
\end{picture}
\end{figure}
\clearpage
\newpage
\begin{figure}[htb]
\setlength{\unitlength}{1cm}
\begin{picture}(15,23)(0,0)
\put(-0.5,10.0){\psfig{figure=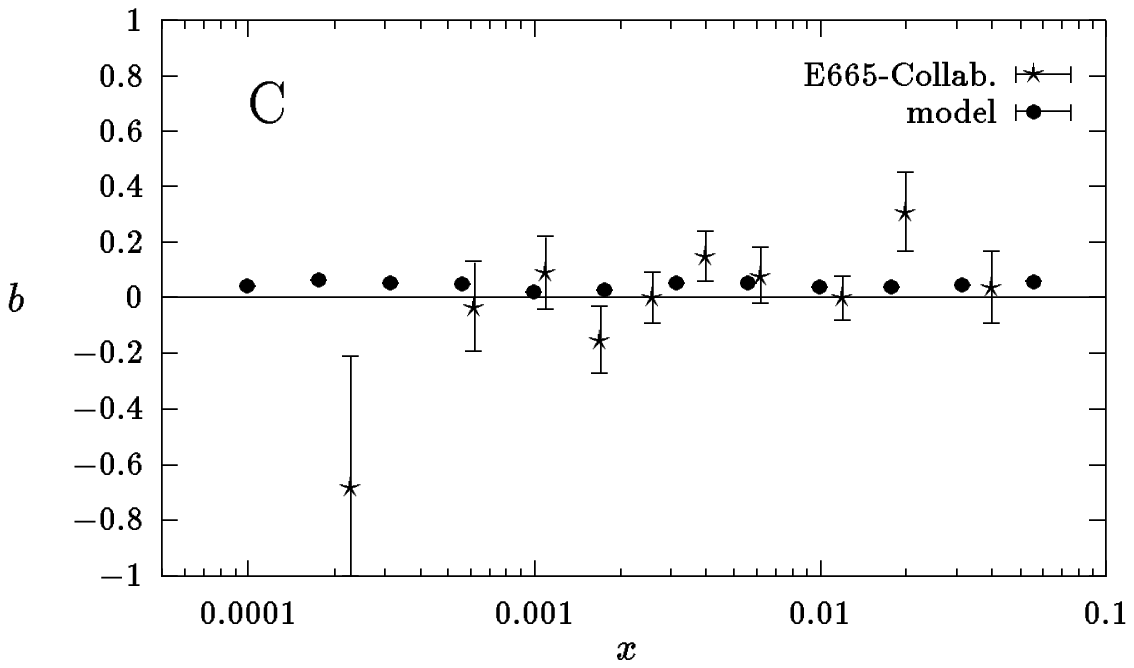}}
\put(3.0,16.0){\bf\large a)}
\put(-0.5,2.5){\psfig{figure=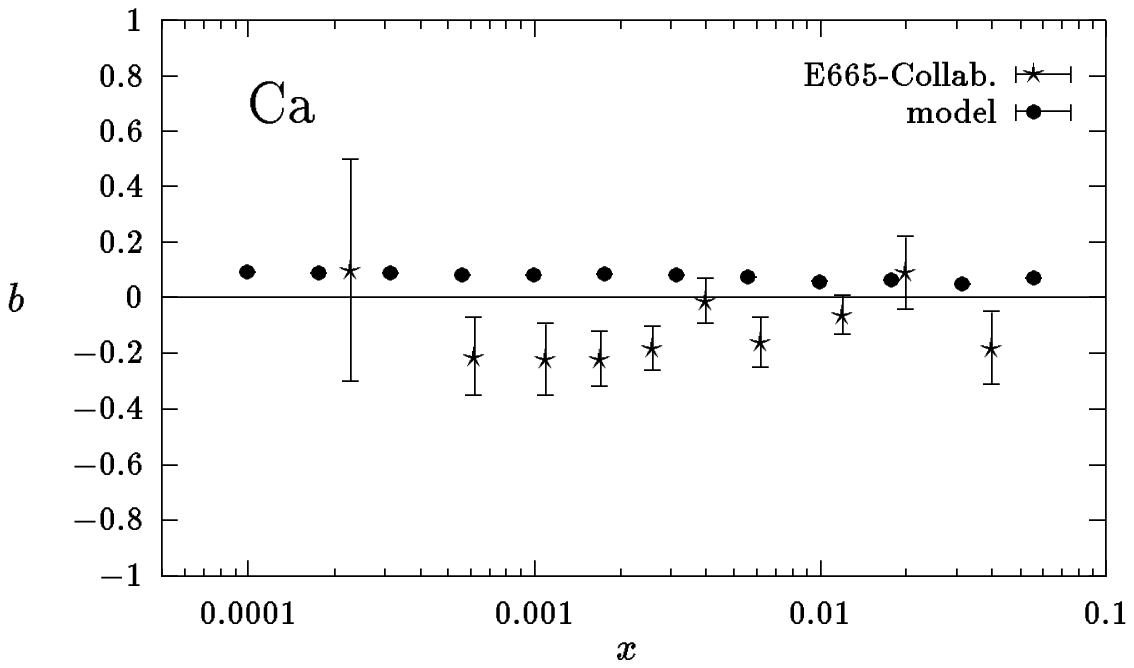}}
\put(3.0,8.5){\bf\large b)}
\put(-0.5,-5.0){\psfig{figure=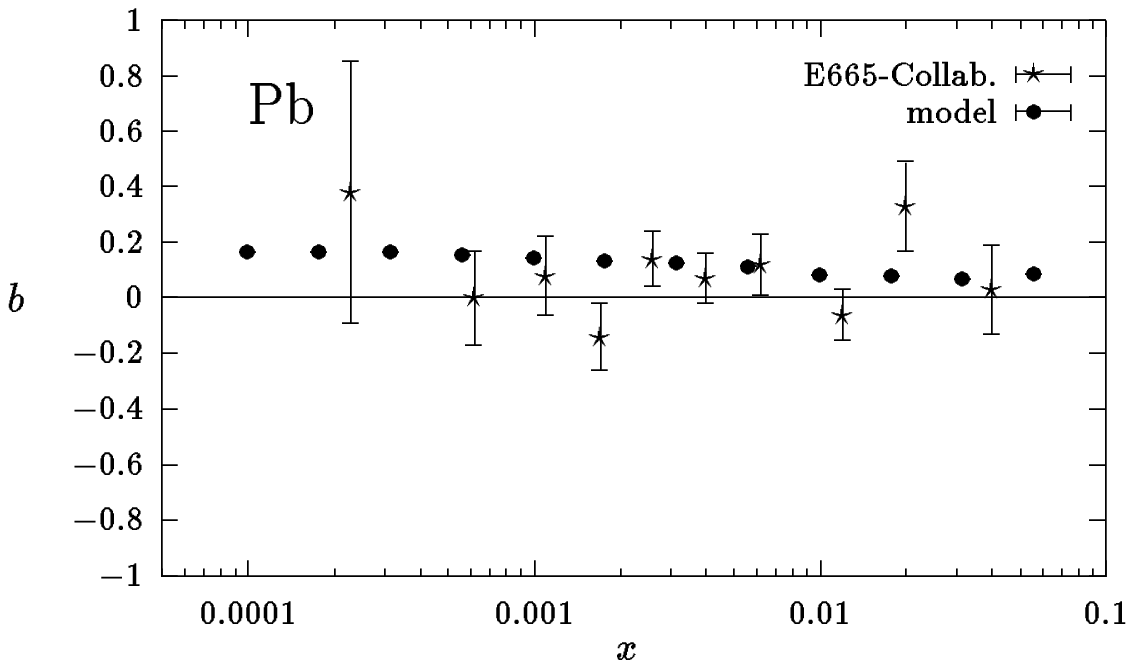}}
\put(3.0,1.0){\bf\large c)}
\put(7.5,0.0){\bf\large Fig.~\ref{gvAsha_b_E665}}
\end{picture}
\end{figure}
\clearpage
\newpage
\begin{figure}[htb]
\setlength{\unitlength}{1cm}
\begin{picture}(15,23)(0,0)
\put(-2.0,0.0){\psfig{figure=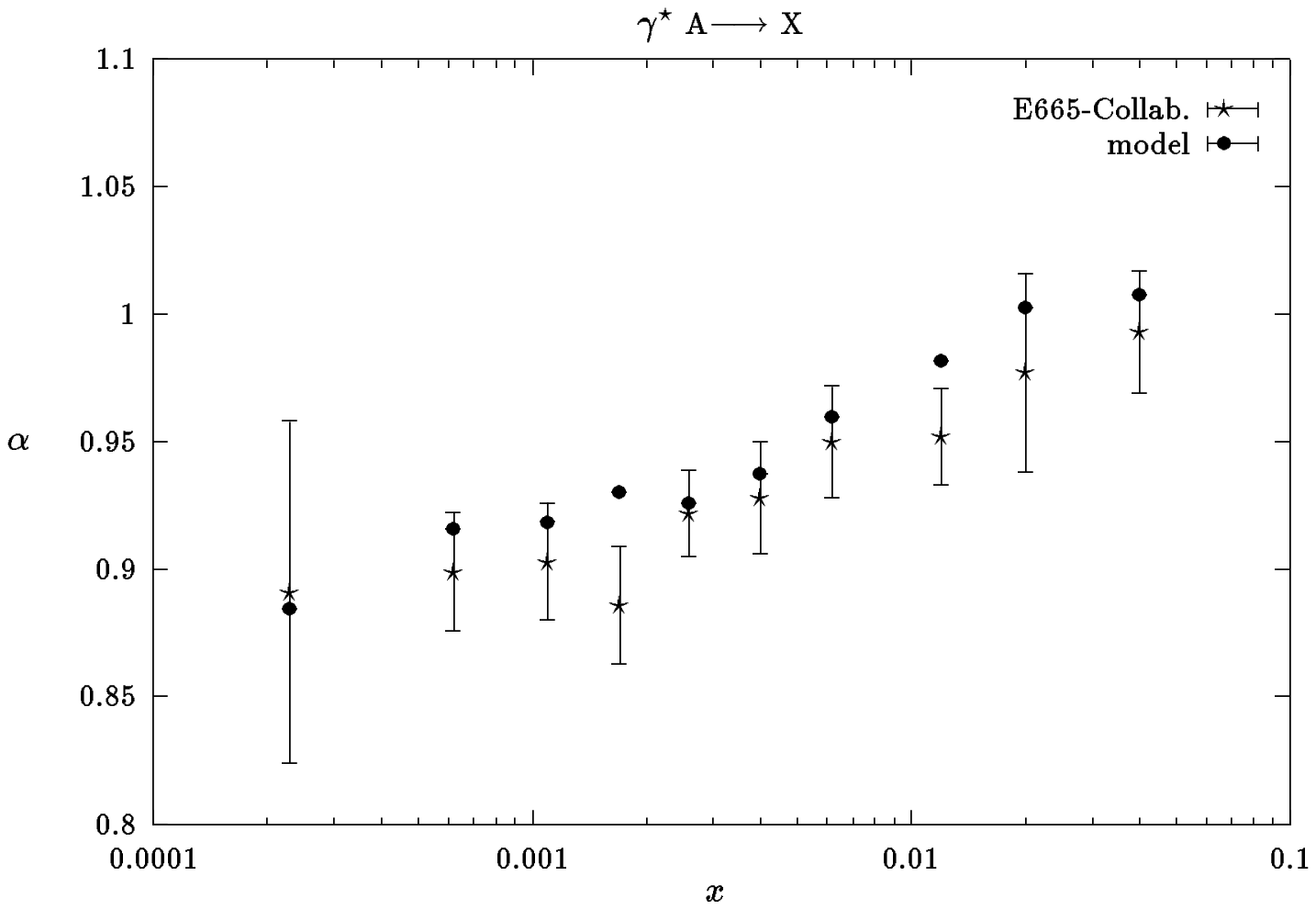}}
\put(7.5,0.0){\bf\large Fig.~\ref{gvAalpha_E665}}
\end{picture}
\end{figure}
\clearpage
\newpage
\begin{figure}[htb]
\setlength{\unitlength}{1cm}
\begin{picture}(15,23)(0,0)
\put(-2.0,7.5){\psfig{figure=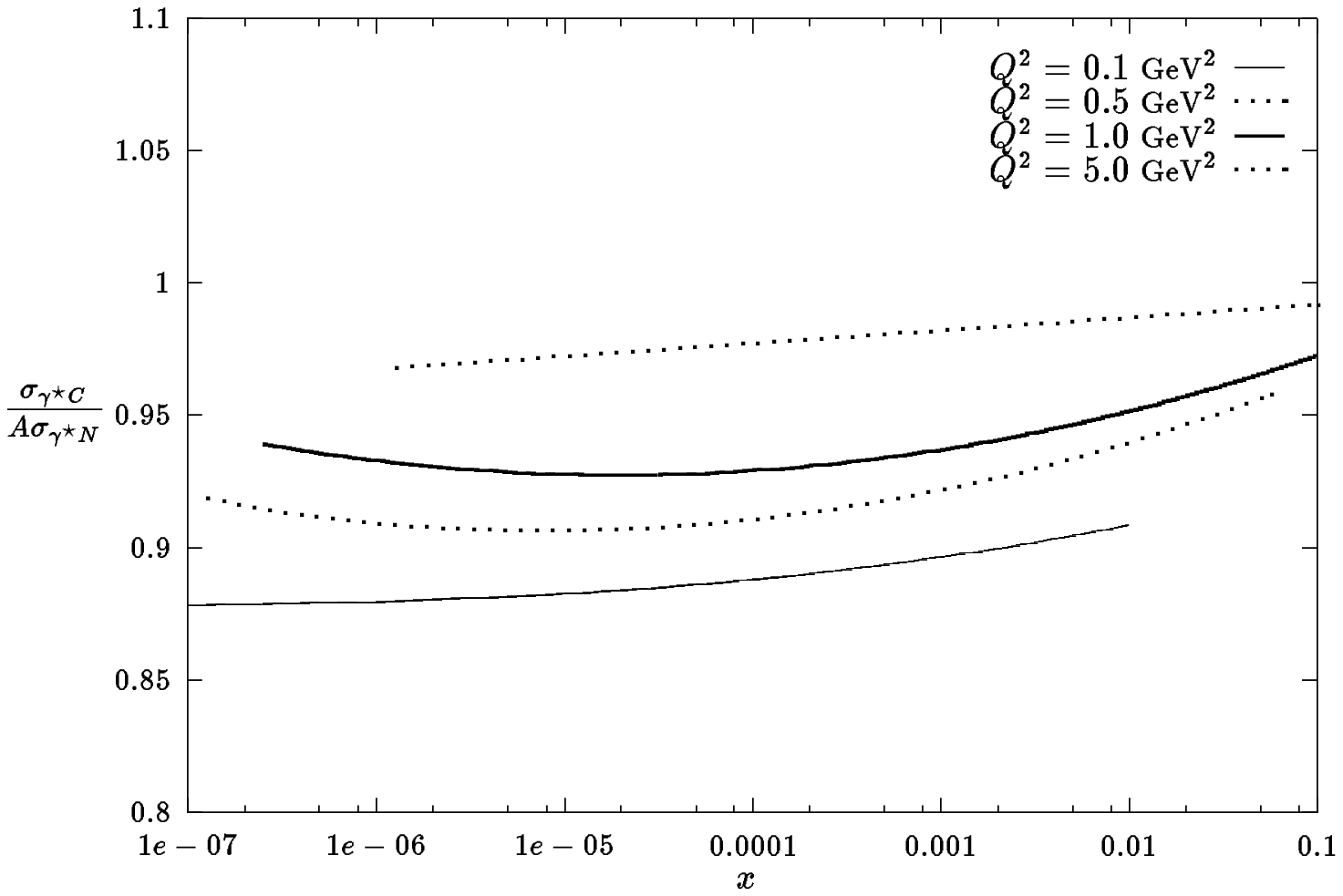}}
\put(0,13.0){\bf\large a)}
\put(-2.0,-4.5){\psfig{figure=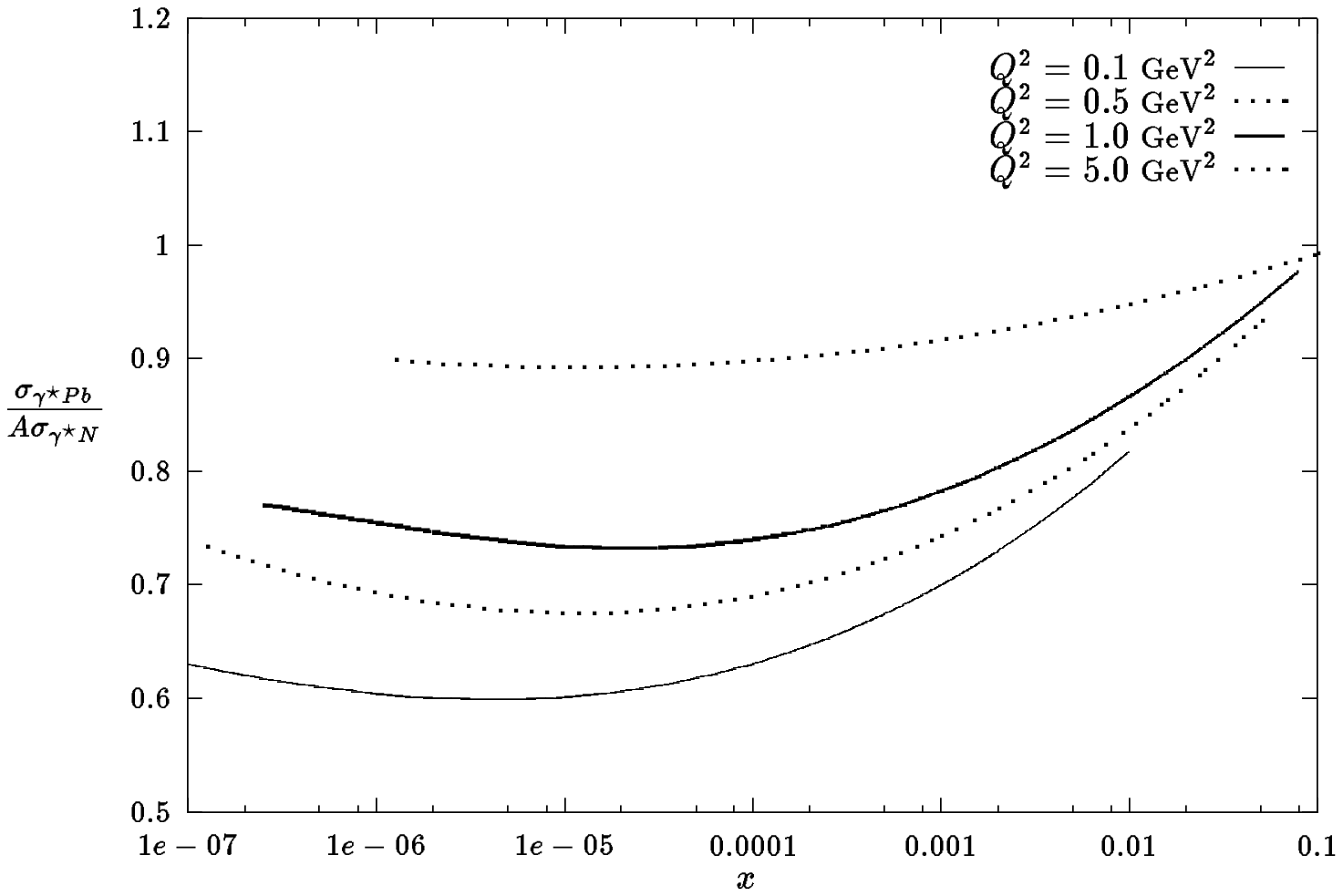}}
\put(0,1.0){\bf\large b)}
\put(7.5,0.0){\bf\large Fig.~\ref{gvAsha_kop}}
\end{picture}
\end{figure}
\clearpage
\newpage
\begin{figure}[htb]
\setlength{\unitlength}{1cm}
\begin{picture}(15,23)(0,0)
\put(-0.5,10.0){\psfig{figure=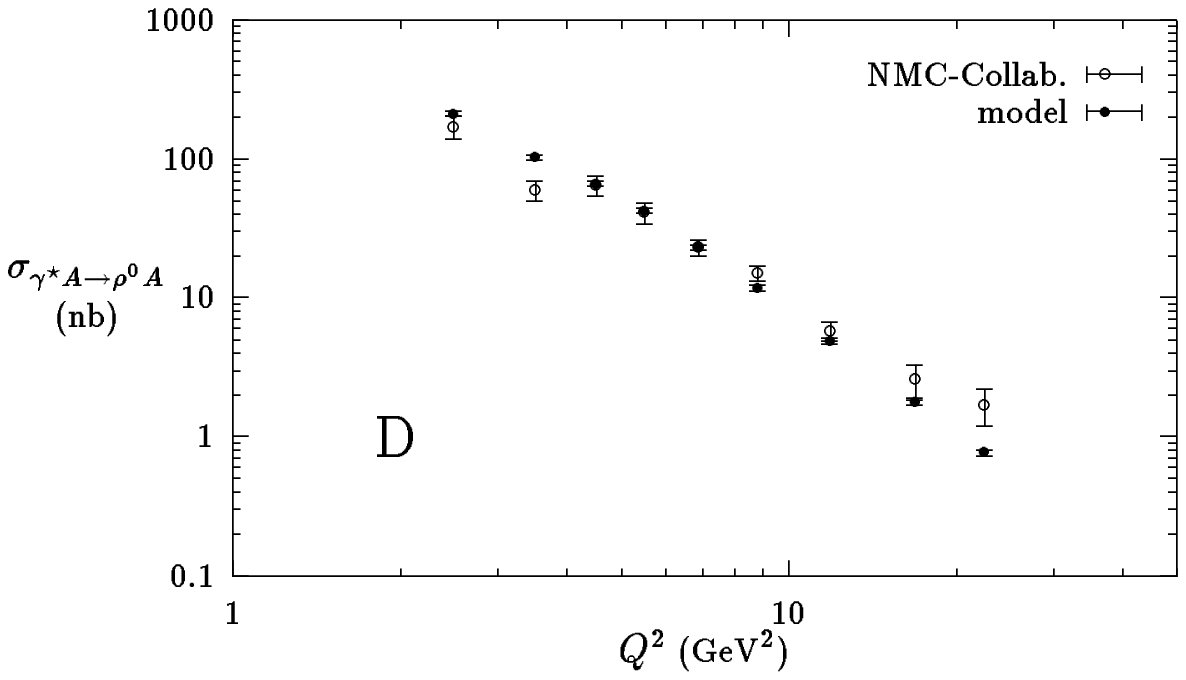}}
\put(3.0,16.0){\bf\large a)}
\put(-0.5,2.5){\psfig{figure=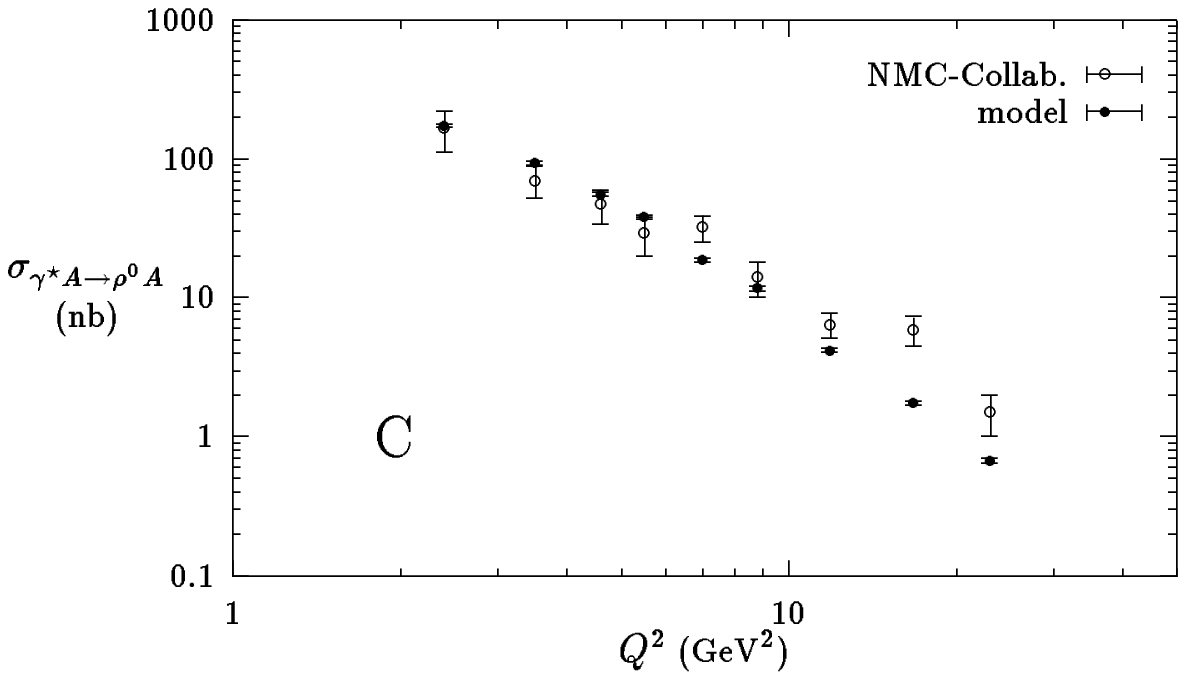}}
\put(3.0,8.5){\bf\large b)}
\put(-0.5,-5.0){\psfig{figure=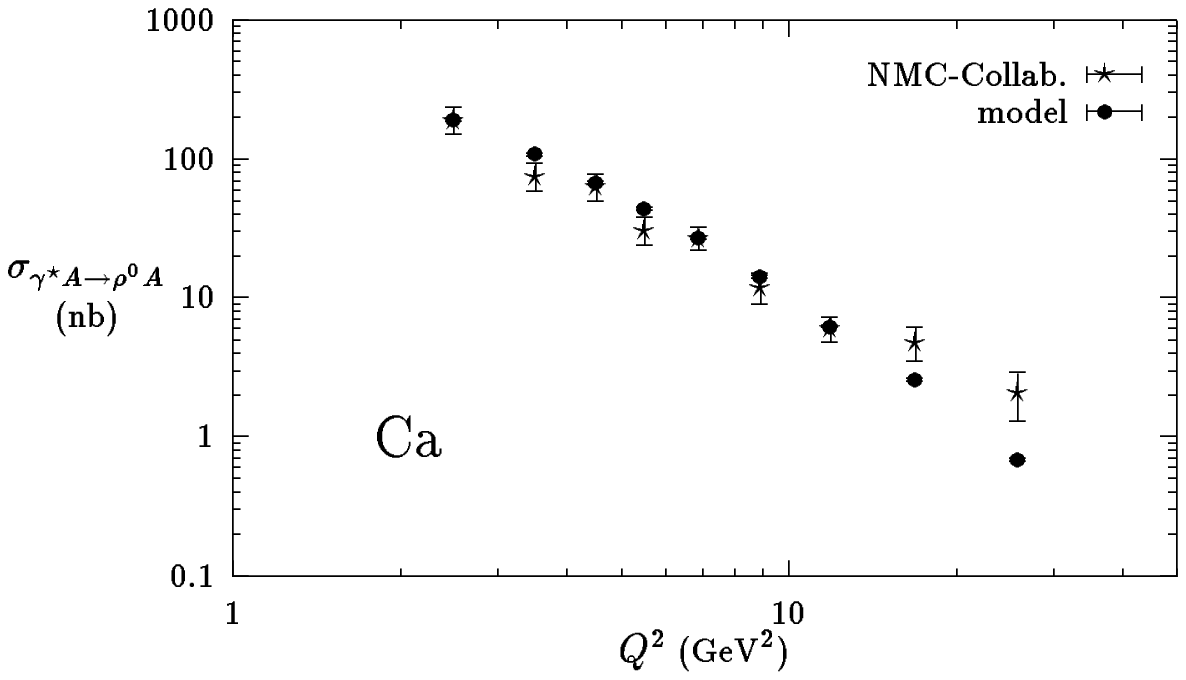}}
\put(3.0,1.0){\bf\large c)}
\put(7.5,0.0){\bf\large Fig.~\ref{gAqelrho}}
\end{picture}
\end{figure}

\clearpage
\begin{thebibliography}{10}

\bibitem{Erdmann96}
M.~Erdmann:
\newblock {The Partonic Structure of the Photon: Photoproduction at the
  Lepton-Proton Collider HERA},
\newblock DESY 96-090,
\newblock 1996

\bibitem{Ahmed92a}
H1 Collab.:  T.~Ahmed et~al.:
\newblock Phys.\ Lett.\ B297 (1992) 205

\bibitem{Derrick94a}
ZEUS Collab.:  M.~Derrick et~al.:
\newblock Phys.\ Lett.\ B322 (1994) 287

\bibitem{Aid95c}
H1 Collab.:  S.~Aid et~al.:
\newblock Z.\ Phys.\ C70 (1995) 17

\bibitem{Derrick95f}
ZEUS Collab.:  M.~Derrick et~al.:
\newblock Phys.\ Lett.\ B354 (1995) 136

\bibitem{Bauer78}
T.~H. Bauer, R.~D. Spital  and D.~R. Yennie:
\newblock Rev.\ Mod.\ Phys.\ 50 (1978) 261

\bibitem{Donnachie78}
G.~{Grammer, Jr.} and D.~Sullivan:
\newblock {\em {Nuclear Shadowing of Electromagnetic Processes,}}
\newblock Plenum Press, New York, 1978
\newblock {in: Electromagnetic Interactions of Hadrons, Volume 2, ed.~by A.\
  Donnachie and G.\ Shaw}

\bibitem{Engel95a}
R.~Engel:
\newblock Z.\ Phys.\ C66 (1995) 203

\bibitem{Engel95d}
R.~Engel and J.~Ranft:
\newblock Phys.\ Rev.\ D54 (1996) 4244

\bibitem{Witten77}
E.~Witten:
\newblock Nucl.\ Phys.\ B120 (1977) 189

\bibitem{Arneodo94b}
M.~Arneodo:
\newblock Phys.\ Rep.\ 240 (1994) 301

\bibitem{Piller95}
G.~Piller, W.~Ratzka  and W.~Weise:
\newblock Z.\ Phys.\ A352 (1995) 427

\bibitem{Glauber55}
R.~Glauber:
\newblock Phys.\ Rev.\ 100 (1955) 242

\bibitem{Gribov70}
V.~N. Gribov:
\newblock JETP 30 (1970) 709

\bibitem{Bertocchi72}
L.~Bertochi:
\newblock Il Nuovo Cimento 11A (1972) 45

\bibitem{Frankfurt89}
L.~L. Frankfurt and M.~I. Strikman:
\newblock Nucl.\ Phys.\ B316 (1989) 340

\bibitem{Frankfurt94}
L.~L. Frankfurt, G.~A. Miller  and M.~I. Strikman:
\newblock Annu.\ Rev.\ Nucl.\ Part.\ Sci.\ 45 (1994) 501

\bibitem{Davidenko78}
G.~V. Davidenko and N.~N. Nikolaev:
\newblock Nucl.\ Phys.\ B135 (1978) 333

\bibitem{Ioffe84a}
B.~L. Ioffe, V.~A. Khoze  and L.~N. Lipatov:
\newblock {\em {Hard Processes, Volume 1: Phenomenology Quark-Parton Model,}}
\newblock North-Holland Physics Publishing, Amsterdam, 1984

\bibitem{Huefner96a}
J.~H\"ufner, B.~Kopeliovich  and J.~Nemchik:
\newblock Phys.\ Lett.\ B383 (1996) 362

\bibitem{Engel96c}
R.~Engel, J.~Ranft  and S.~Roesler:
\newblock {Photoproduction off nuclei and point-like photon interactions; 
  Part II: Particle production},
\newblock in preparation,  1996

\bibitem{Sakurai72a}
J.~J. Sakurai and D.~Schildknecht:
\newblock Phys.\ Lett.\ 40B (1972) 121

\bibitem{Gorczyca73}
B.~Gorczyca and D.~Schildknecht:
\newblock Phys.\ Lett.\ 47B (1973) 71

\bibitem{Budnev75}
V.~M. Budnev, I.~F. Ginzburg, G.~V. Meledin  and V.~G. Serbo:
\newblock Phys.\ Rep.\ 15C (1975) 181

\bibitem{Capella94b}
A.~Capella, A.~Kaidalov, C.~Merino  and J.~Tran Thanh~Van:
\newblock Phys.\ Lett.\ B337 (1994) 358

\bibitem{Abramowicz91a}
H.~Abramowicz, E.~M. Levin, A.~Levy  and U.~Maor:
\newblock Phys.\ Lett.\ B269 (1991) 465

\bibitem{Badelek92}
B.~Badelek and J.~Kwieci\'nski:
\newblock Phys.\ Lett.\ B295 (1992) 263

\bibitem{Alekhin87}
S.~I. Alekhin et~al.:
\newblock Compilation of cross sections 4,
\newblock CERN-HERA 87-01,
\newblock 1987

\bibitem{Derrick94b}
ZEUS Collab.:  M.~Derrick et~al.:
\newblock Z.\ Phys.\ C63 (1994) 391

\bibitem{Aid95b}
H1 Collab.:  S.~Aid et~al.:
\newblock Z.\ Phys.\ C69 (1995) 27

\bibitem{Schuler93b}
G.~A. Schuler and T.~Sj\"ostrand:
\newblock Nucl.\ Phys.\ B407 (1993) 539

\bibitem{Sjostrand85}
T.~Sj\"ostrand:
\newblock Phys.\ Lett.\ B157 (1985) 321

\bibitem{Gottschalk86}
T.~Gottschalk:
\newblock Nucl.\ Phys.\ B277 (1986) 700

\bibitem{GRV92b}
M.~Gl\"uck, E.~Reya  and A.~Vogt:
\newblock Phys.\ Rev.\ D46 (1992) 1973

\bibitem{GRV92a}
M.~Gl\"uck, E.~Reya  and A.~Vogt:
\newblock Phys.\ Rev.\ D45 (1992) 3986

\bibitem{Shmakov89}
S.~Y. Shmakov, V.~V. Uzhinskii  and A.~M. Zadoroshny:
\newblock Comp.\ Phys.\ Commun.\ 54 (1989) 125

\bibitem{Haakman96}
L.~P.~A. Haakman, A.~Kaidalov  and J.~H. Koch:
\newblock Phys.\ Lett.\ B365 (1996) 411

\bibitem{Segre77}
E.~Segr\'e:
\newblock {\em Nuclei and particles}
\newblock Reading Mass Benjamin 1977

\bibitem{Kopeliovich96b}
B.~Kopeliovich:
\newblock {Soft Component of Hard Reactions and Nuclear Shadowing},
\newblock {in Proceedings of the Workshop Hirschegg'95: Dynamical 
  Properties of
  Hadrons in Nuclear Matter, ed.~by H.\ Feldmeier and W.\ N\"orenberg,
  Darmstadt, p.\ 102},  1995

\bibitem{Arakelyan78}
E.~A. Arakelyan et~al.:
\newblock Phys.\ Lett.\ 79B (1978) 143

\bibitem{Brookes73}
G.~R. Brookes et~al.:
\newblock Phys.\ Rev.\ D8 (1973) 2826

\bibitem{Caldwell73}
D.~O. Caldwell et~al.:
\newblock Phys.\ Rev.\ D7 (1973) 1362

\bibitem{Caldwell79}
D.~O. Caldwell et~al.:
\newblock Phys.\ Rev.\ Lett.\ 42 (1979) 553

\bibitem{Adams95}
E665-Collab.:  M.~R. Adams et~al.:
\newblock Z.\ Phys.\ C67 (1995) 403

\bibitem{Amaudruz91}
New Muon Collab.:  P.~Amaudruz et~al.:
\newblock Z.\ Phys.\ C51 (1991) 387

\bibitem{Arneodo95}
New Muon Collab.:  M.~Arneodo et~al.:
\newblock Nucl.\ Phys.\ B441 (1995) 12

\bibitem{Kopeliovich96a}
B.~Kopeliovich and B.~Povh:
\newblock {Interplay of Soft and Hard Interactions in Nuclear Shadowing at 
  High $Q^2$ and Low $x$},
\newblock MPI H-V28-1996,
\newblock 1996

\bibitem{Geiger93}
K.~Geiger:
\newblock Phys.\ Rev.\ D47 (1993) 133

\bibitem{Bopp94a}
F.~W. Bopp, R.~Engel, D.~Pertermann  and Ranft:
\newblock Phys.\ Rev.\ D49 (1994) 3236

\bibitem{Arneodo94a}
New Muon Collab.:  M.~Arneodo et~al.:
\newblock Nucl.\ Phys.\ B429 (1994) 503

\end{thebibliography}
\end{document}